\documentclass[aps,prd,twocolumn,nofootinbib,floatfix]{revtex4}
\usepackage{graphicx}
\usepackage[utf8]{inputenc}
\usepackage{placeins}
\usepackage{amssymb, amsmath}
\usepackage{tabularx}
\newcommand{\Ms}{M_{\odot}}

\newcommand{\rns}{\rho_{\rm sat}}

\newcommand{\dops}{doppelg\"angers~}

\newcommand{\dop}{doppelg\"anger~}

\usepackage[dvipsnames, usenames]{xcolor}

\begin{document}

\title{Tidal Deformability Doppelg\"angers:\\
\textit{Implications of a low-density phase transition in the neutron star equation of state}}
\author{Carolyn A. Raithel}
\email{craithel@ias.edu}
\author{Elias R. Most}
\email{emost@princeton.edu}
\affiliation{School of Natural Sciences, Institute for Advanced Study, 1 Einstein Drive, Princeton, NJ 08540, USA}
\affiliation{Princeton Center for Theoretical Science, Jadwin Hall, Princeton University, Princeton, NJ 08544, USA}
\affiliation{Princeton Gravity Initiative, Jadwin Hall, Princeton University, Princeton, NJ 08544, USA}
\date{August 2022}

\begin{abstract}
Studying the properties of ultra-dense matter is one of the key goals of
modern neutron star research. The measurement of the tidal deformability
from the inspiral of a binary neutron star merger offers one promising
method for constraining the equation of state (EoS) of cold, dense matter.
In this work, we report on a new class of EoSs
which have significantly different pressures at nuclear densities
and large differences in stellar radii, but that
predict surprisingly similar tidal deformabilities across the
entire range of astrophysically-observed neutron star masses. 
Using a survey of 5 million piecewise polytropic EoSs, subject to
five different sets of nuclear priors, we demonstrate that these 
``tidal deformability \dops" occur
generically. We find that they can differ
substantially in the pressure (by up to a factor of 3 at nuclear densities)
and in the radius of intermediate-mass neutron stars (by up to 0.5~km),
but are observationally indistinguishable in their tidal deformabilities
($\Delta\Lambda \lesssim 30$) with the sensitivity of 
current gravitational wave detectors. We demonstrate that this near-degeneracy in the
tidal deformability is a result of allowing for a phase transition at low
densities. We show that a combination of
input from nuclear theory (e.g.,
from chiral effective field theory), X-ray observations of neutron star radii,
and/or the next generation of gravitational wave detectors
will be able to significantly constrain these tidal deformability
\dops.

\end{abstract}

\maketitle

\section{Introduction}

Astrophysical observations of neutron stars provide a unique laboratory
for constraining the equation of state (EoS) of ultra-dense matter. Such
constraints have been made, for example, using
measurements of the neutron star radius
from spectral modeling of bursting or
quiescent neutron stars in X-ray binaries \cite{Ozel2009a,Guver2010,Guillot2013,Guillot2014,Heinke2014,Nattila2015,Ozel2016a,Bogdanov2016,Ozel2016}
or, more recently,
from pulse-profile modeling of X-ray pulsars with the Neutron star
Interior Composition ExploreR (NICER)
\cite{Miller:2019cac,Riley:2019yda,Miller:2021qha,Riley:2021pdl}.
With the advent of gravitational wave (GW) astronomy, 
a complementary avenue for constraining the EoS is now also possible.

In particular, observations of the inspiral gravitational
waves from the first binary neutron star merger, GW170817, have
constrained the tidal deformability of a 1.4$\Ms$ star to $\Lambda_{1.4}=190\substack{+390\\-120}$ at 90\% confidence,
which in turn has been used to constrain
the pressure at twice the nuclear saturation density to within
$\sim125$\% \cite{LIGOScientific:2017vwq,LIGOScientific:2018hze,LIGOScientific:2018cki}.
It is projected that within the next five years, the LIGO-Virgo-Kagra
network is likely to detect $\mathcal{O}(10-20)$ additional neutron star mergers with
high signal-to-noise ratios (SNR), 
which could further constrain the pressure at twice nuclear densities to within a factor of $\sim2$
\cite{Lackey2015,Essick:2019ldf,Landry:2020vaw,HernandezVivanco:2020cyp}; or, potentially to within 20\% 
given certain assumptions about the nuclear EoS \cite{Forbes:2019xaz}.

With the construction of next-generation GW detectors such as 
Cosmic Explorer \cite{Reitze:2019iox}, Einstein Telescope \cite{Punturo:2010zz},
or NEMO \cite{Ackley:2020atn}, even tighter
constraints on the EoS will become possible in the 2030s. For example, an event like
GW170817 would have an SNR of 2800 with Cosmic Explorer,
roughly 88 times larger than was observed in 2017  \cite{Carson:2019rjx}.
These detectors will observe tens of thousands 
of events per year, hundreds of which will have extremely high
SNR in the inspiral, 
leading to anticipated constraints on the binary tidal deformability of
$\sigma_{\widetilde{\Lambda}} < 20$ and thus
enabling a new era of precision EoS constraints \cite{Chatziioannou:2021tdi}.

The prospects for constraining the EoS with current or next-generation GW
observations relies on the unique mapping between the tidal deformability and
the underlying EoS. In the standard paradigm, these constraints would be 
limited only by the sensitivity to which the tidal deformability
can be measured, and the masses at which it is measured. In practice,
effects such as dynamical tides \cite{Pratten:2021pro,Gamba:2022mgx}
or systematics in the available waveform models \cite{Gamba:2020wgg} may
complicate the extraction of the tidal deformability from the inspiral GWs,
but the tidal deformability itself is assumed to map robustly to the EoS at supranuclear densities.

In this paper, we describe a new construction of EoS models that poses
a challenge to this paradigm.  In particular, we
demonstrate that uncertainties in the EoS above nuclear densities lead
to the emergence of what we call ``tidal deformability doppelg\"angers":  these are 
EoS models that differ significantly in pressure at supranuclear
densities and accordingly in the neutron star radius, but that predict 
nearly identical tidal deformabilities across a wide range of neutron star masses.
The most extreme of these \dop EoSs can vary in the pressure by factors of $\sim$3
and in the radius by up to 0.5~km, 
but they differ in the tidal deformability by $\lesssim10$ across the entire range
of astrophysically-observed neutron star masses,
making them observationally indistinguishable to current GW detectors.

We find that pairs of \dop EoSs are ubiquitous in randomly-generated
EoS samples and that they occur as a natural consequence of allowing for a phase transition
in the EoS at densities between 1 and 2 times the nuclear saturation density
(where the exact density depends on the details of the crust EoS
and the high-density parameterization, as we will demonstrate).
We demonstrate that these \dops can be constructed by exploiting the different density
dependencies of the tidal Love number, $k_2$, and the stellar compactness, $C$,
such that the tidal deformability $\Lambda = (2/3) k_2 C^{-5}$ remains
the same, in spite of large differences in the stellar radii 
and pressures at supranuclear densities. Interestingly, 
we find that the \dops approximately obey the quasi-universal relations between the tidal deformability and the
moment of inertia, but that the \dops tend to fall below the standard quasi-universal relations with stellar compactness \cite{Yagi:2016bkt}.

The small differences in the tidal deformability curves of the \dop models 
will likely require the
next-generation of GW detectors to resolve. We demonstrate, however, that
the parameter space that is subject to this tidal
deformability degeneracy can also be reduced
by applying more restrictive nuclear priors, e.g. on the crust EoS or on the density-derivatives of the pressure.
At low-densities, such constraints may come from
nuclear theory (e.g., from chiral EFT \cite{Gezerlis:2013ipa,Lynn:2015jua,Tews:2015ufa,Drischler:2017wtt}); while new measurements of
neutron star radii 
 or even broad constraints on the
tidal deformability with current GW detectors may also help to constrain
the parameter space of the the tidal deformability \dops,
even before the advent of next-generation detectors.

The outline of the paper is as follows. In Sec.~\ref{sec:intro_dopps}, we introduce several examples of
\dop EoSs and illustrate their basic construction. In Sec.~\ref{sec:param}, we perform a large-scale EoS parameter
survey with different sets of nuclear priors, to characterize the regions 
of EoS parameter space where the tidal-deformability degeneracy can occur and we quantify
the unique
signatures of these models. In Sec.~\ref{sec:discussion},
we discuss the implications of the \dop EoSs and the prospects for resolving the degeneracy with
joint input from nuclear physics and astrophysics.

\section{Observational degeneracy between EoS models with low-density phase transitions}
\label{sec:intro_dopps}

We start in this section by introducing a few examples of tidal deformability \dops,
in order to illustrate their key features.

Here and throughout this paper, we utilize parametric models of the EoS to explore the EoS parameter space in search
of \dop models. In particular, we adopt a piecewise polytropic (PWP) parametrization of the EoS, using
five polytropic segments that are spaced uniformly in the logarithm of the density
between $\rho_0$ and 7.4$\rns$ \cite{Read2009,Ozel2009,Raithel2016},
where $\rns=2.7\times 10^{14}$g/cm$^3$ is the nuclear saturation density. The
starting density of the parametrization is taken to be 
between $\rns$ and $1.5\rns$, below which we adopt a tabular, nuclear EoS
to describe the crust. For these crust EoSs, we choose nuclear models that are
consistent with low-density nuclear constraints. We
fix the pressure at $\rho_0$ to that of the crust EoS, in order to 
ensure continuity in the resulting model. The five polytropic segments
of the PWP EoS
are then determined by specifying the pressures at each fiducial density
in the model.

In constructing
new EoSs, we impose a set of minimal physical constraints
on the EoS above $\rho_0$, namely that:
\begin{enumerate}
    \item The EoS must be hydrostatically stable (i.e., $\partial P / \partial \rho \ge 0$, where $P$ is the pressure and $\rho$ is the rest-mass density).
    \item The sound speed must remain subluminal at all densities.
    \item The EoS must be consistent with the observation of massive pulsars. In particular, we require that $M_{\rm max} > 2.01 \Ms$,
    which corresponds to the $1\sigma$ lower limit on the current most massive neutron star \cite{NANOGrav:2019jur,Fonseca:2021wxt}. 
\end{enumerate}
Once these minimal constraints are satisfied, we exploit the uncertainties
in the EoS to freely explore the remaining parameter space. We note that the resulting EoSs are quite broad in their coverage.
We intentionally explore the extremes of the EoS parameter space in order to identify and highlight the new behavior of the \dop models. As we will demonstrate, by applying additional nuclear input, the ubiquity and extremity of the \dops can be significantly reduced.

 \begin{figure*}[!ht]
\centering
\includegraphics[width=0.8\textwidth]{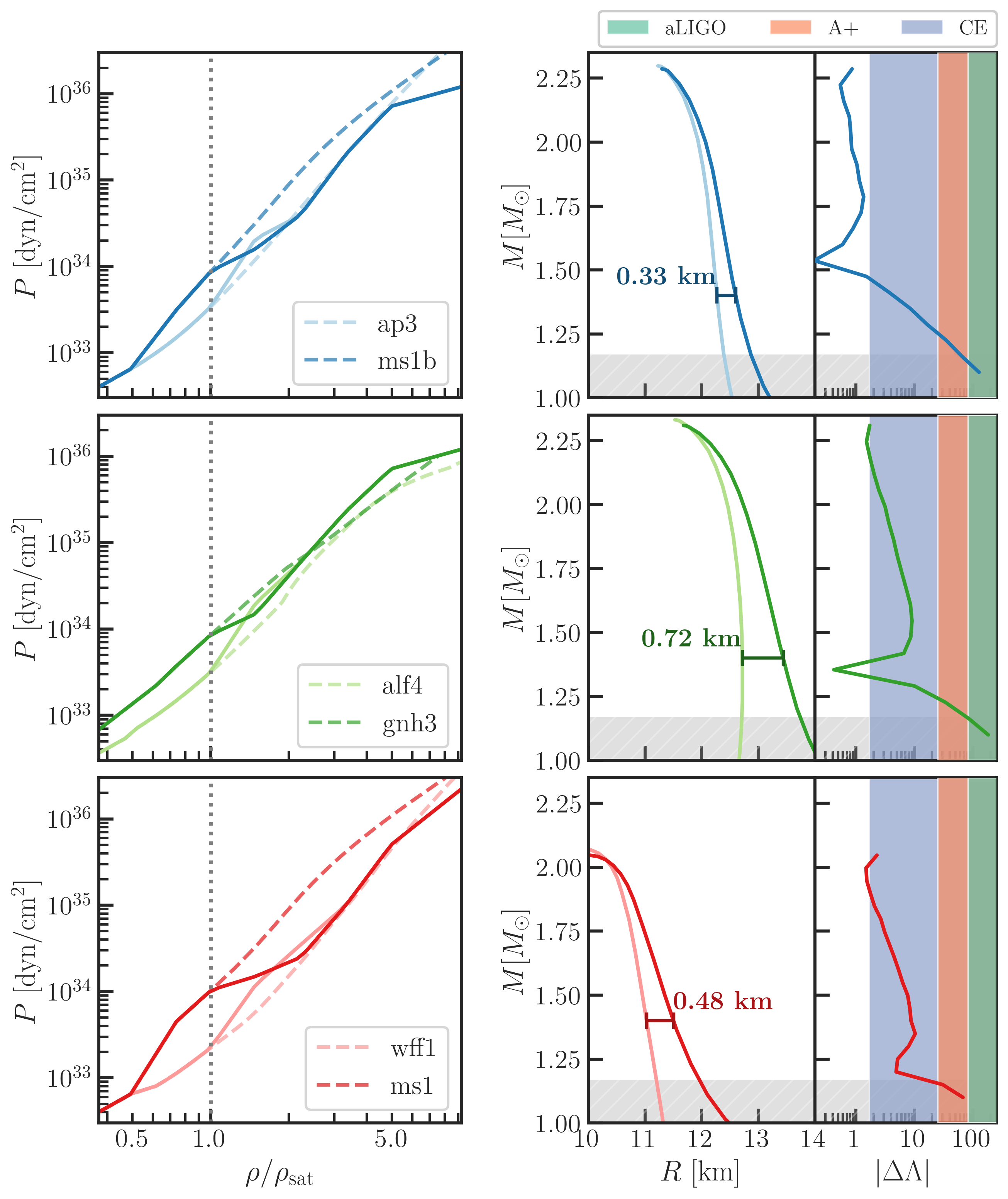}
\caption{\label{fig:tabEOSs} Three example pairs of \dops, each constructed
by assuming a different microphysical EoS for the crust. The various crust
EoSs (which are shown with dashed lines) are used up to $\rns$; at higher
densities, we use piecewise polytropes to explore the parameter space and
to illustrate the degeneracy in tidal deformabilities. The EoSs are shown
in the left column; the middle column shows the corresponding mass-radius
relations; and the right column shows the absolute difference in tidal
deformability at each mass, for a given pair of \dops. The vertical bands indicate the expected 68\%-measurement uncertainty in the tidal deformability for a population of neutron star
  mergers, assuming an intermediate merger detection rate, for 
  the sensitivity of LIGO at design sensitivity (aLIGO), 
  the anticipated sensitivity of LIGO during its fifth observing run (A+),
  and the proposed XG detector Cosmic Explorer (CE)
  \cite{Carson:2019rjx}. Masses below the lightest-observed
  neutron star mass are masked in gray. }
\end{figure*}

 \begin{figure*}[!ht]
\centering
\includegraphics[width=\textwidth]{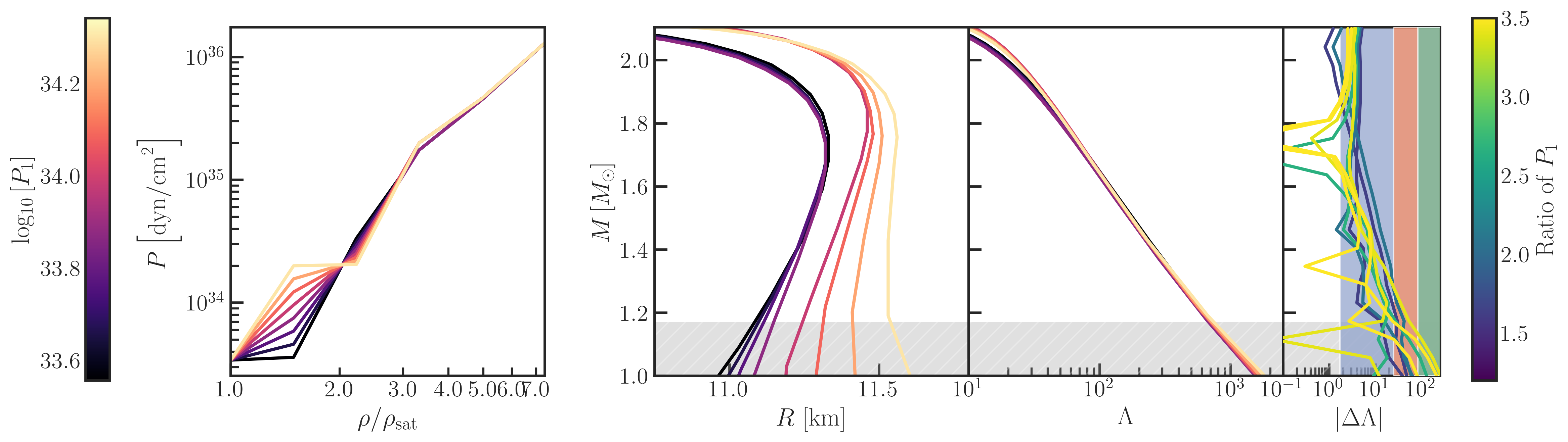}
\caption{\label{fig:degeneracy} Tidal deformability degeneracy for an
example family of \dop EoSs. From left to right, we show: the EoS pressure
as a function of density, the mass-radius ($M-R$) relation for each EoS,
and the mass-tidal deformability relation. The curves in these
first three figures are colored according to the EoS pressure at 
the first fiducial density, $1.5\rns$, where $\rns$ is the nuclear saturation density.
The figure on the far right shows the absolute difference in $\Lambda$ between
any two of these EoSs, with a blue-green color scale to indicate the ratio
of pressures at $\rns$ for the two EoSs being compared. The vertical
shaded bands indicate the expected detection sensitivity in $\Lambda$ for a population
of mergers observed with current and next-generation detectors,
and masses below the lightest-observed
  neutron star mass are masked in gray
  (as in Fig.~\ref{fig:tabEOSs}).
For any pair EoSs, the tidal deformabilities differ by $\lesssim$50 across 
the entire range of astrophysical neutron star masses, even
though the pressures at $1.5 \rns$ differ by a factor of 5}.
\end{figure*}

We introduce the concept of tidal-deformability \dops with a few examples in
Fig.~\ref{fig:tabEOSs}. In this first illustration,
we have chosen commonly-used low-density EoSs, onto which we attach
different high-density EoSs that are constructed using PWPs to
produce similar tidal deformability curves. For example, in
the top row of Fig.~\ref{fig:tabEOSs}, the light blue model uses the
variational-method nuclear EoS ap3 \cite{Akmal:1998cf} to describe the crust up to $\rns$;
while the dark blue model uses the relativistic mean-field EoS ms1b
\cite{Mueller:1996pm} for the crust. At $\rho>\rns$, we switch to the PWP
construction described above, and we adjust the high-density pressures to
specifically minimize differences in the tidal deformability $ \Lambda$.
The net result is a pair of EoSs
that differ in the pressure at $\rns$ by a factor of 2.6, in the radius $R_{1.4}$ of a
1.4~$\Ms$ star by 0.33~km, but differ by $\Delta \Lambda \lesssim$10 across
the entire range of astrophysically-observed neutron star masses. The
second and third rows of Fig.~\ref{fig:tabEOSs} show two additional pairs
of \dop EoSs, which are constructed using different sets of crust EoSs:  
gnh3 \cite{Glendenning:1984jr}, alf4 \cite{Alford:2004pf}, wff1 \cite{Wiringa:1988tp}, and ms1, which is
identical to ms1b but features a higher symmetry energy (for additional details
on these models, see e.g. \cite{Read:2008iy}). In Fig.~\ref{fig:tabEOSs}, we have grayed out
masses below 1.17 $\Ms$, which is the 1-$\sigma$ lower limit on
the lightest-observed pulsar \cite{Martinez:2015mya}, in order to focus on
astrophysically-observed neutron star masses.
\footnote{We note that the 90\% lower-limit on the secondary
mass from GW170817 was also 1.17$\Ms$ and that, to-date,
no lighter neutron stars have yet been detected from gravitational wave
events \cite{LIGOScientific:2017vwq}.} In these examples, we
find fractional differences in pressure of 170\% and in $R_{1.4}$ of 0.72~km for the
gnh3/alf4 pair of \dops (in green); and 365\% in pressure and 0.48~km in
$R_{1.4}$ for the wff1/ms1 pair of \dops (in red). 

Despite these large differences in pressure and radius, the tidal
deformabilities for each set of models are very similar in all three cases.
We show the difference in tidal deformability as a function of the mass for
each pair of models in the far right column of Fig.~\ref{fig:tabEOSs}. In
these figures, the vertical shaded bands indicate 
anticipated 68\% measurement uncertainties
in the tidal deformability, $\sigma_{\Lambda}$, for a population
of neutron star mergers, assuming an intermediate merger detection rate
observed with current and next-generation facilities \cite{Carson:2019rjx}.
\footnote{We note that the 
  $\sigma_{\Lambda}$ from \cite{Carson:2019rjx} was calculated for the binary
  tidal deformability, $\widetilde{\Lambda}$. In this paper, we assume equal
  mass binaries for simplicity, in which case $\widetilde{\Lambda}$ reduces to
  the tidal deformability of either star. We present the difference in
  $\widetilde{\Lambda}$ as a function of the mass ratio for two example pairs
  of \dop EoSs in a companion paper \cite{Raithel:2022efm}.} When this
  measurement uncertainty, $\sigma_{\Lambda}$, becomes comparable to the
  intrinsic difference between a pair of models, $\Delta\Lambda$, the two
  EoSs can no longer be distinguished. Already, we see that a
  next-generation detector such as Cosmic Explorer would be required to
  distinguish between any of these pairs of \dops based on their tidal
  deformabilities alone. 

We note that we focus on absolute differences in radii and tidal deformabilities,
because this allows for the most direct comparison 
against the observational sensitivity of experiments such as NICER or LIGO. For example, it has been estimated that the sensitivity of next-generation GW detectors will yield constraints on the neutron star radii of 50-200~m \cite{Chatziioannou:2021tdi}, significantly smaller
than what we construct for the \dop models shown in Fig.~\ref{fig:tabEOSs}.
To explore the impact of this tidal deformability degeneracy for current and future GW detector sensitivities, we perform mock Bayesian inferences
of the EoS for a pair of \dop models in a companion Letter \cite{Raithel:2022efm}.
  
    The examples shown here were constructed by using different theoretical
  calculations for the crust EoS up to $\rns$ and freely varying the PWP pressures 
  at higher densities. 
  These results are consistent with previous work that has shown that changing the crust EoS (at densities below $10^{14}$~g/cm$^3$) can affect the radius without significantly changing the tidal deformabilities \cite{Gamba:2019kwu}. Here, however, we find that EoSs can have significant ($\sim3\times$) differences at \textit{supranuclear} densities, and still be indistinguishable in their tidal deformabilities.

We can also see this behavior emerge more generically, by
systematically varying the EoS at densities above $\rns$ to explore
the uncertainties in the
EoS at supranuclear densities. We do so in Fig.~\ref{fig:degeneracy}, where we adopt a
single
low-density EoS, ap3, which we use at densities below $\rns$ for all models. 
At densities above $\rns$, we construct a sequence of models that
exhibit similar tidal deformability curves as the supranuclear pressures are
varied.\\
\\

In Fig.~\ref{fig:degeneracy}, it becomes clear that there is actually a
continuum of EoS models that produce nearly-degenerate tidal deformability
curves. The mass-radius and tidal deformability curves for these models 
are shown in the middle panels
of Fig.~\ref{fig:degeneracy}, where they are colored according to the
pressure at the nuclear saturation density, $P_{\rm sat}$. The far right
panel of Fig.~\ref{fig:degeneracy} shows the difference in tidal
deformability between any two pairs of these EoSs, colored according to the
ratio of pressures at $1.5\rns$. As we saw in Fig.~\ref{fig:tabEOSs}, the
tidal deformabilities of these models are nearly indistinguishable with current
GW detectors for all but the lowest-mass systems. 
In this case, the \dops would be distinguishable in an EoS inference
from GW data only by the choice of informative nuclear priors.
In contrast, for a population
of mergers observed with Cosmic Explorer,
these models start to become distinguishable from the data directly.

The fact that the near-degeneracy in tidal deformability curves applies to a
continuous range of EoS parameter space has important implications: it
suggests that even as the range of EoS pressures is further constrained by
future astrophysical detections and by input from nuclear theory, the tidal
deformability degeneracy will be reduced, but not entirely eliminated. We
revisit this point and discuss practical ways for further resolving the
degeneracy in Sec.~\ref{sec:discussion}.

\section{EoS parameter survey}
\label{sec:param}

Having now introduced the common features of a few examples of
tidal-deformability \dops, in this section, we turn to a large-scale
parameter survey, in order to illustrate the ubiquity of the \dop EoSs and to
quantify the specific regions of
parameter space that are susceptible to this tidal-deformability
degeneracy.  

\subsection{Construction of parameteric EoSs}
\label{sec:construction}

To that end, we construct five samples of piecewise polytropic EoSs. All EoSs
consist of five polytropic segments, 
as described in Sec.~\ref{sec:intro_dopps}.
We consider three fiducial densities for the onset of
our piecewise polytopes,
$\rho_0=\{1,1.2,\text{and }1.5\}\rns$, in order to gauge the impact of this starting density
on the \dop behavior.  At densities below this fiducial value,
we adopt a tabulated, nuclear EoS to describe the crust. 
The pressure at $\rho_0$ is set
by the crust EOS, in order to ensure continuity in the EoS.
We explore two choices for the crust EoS. For the baseline
set of models, we use the nuclear model ap3
\cite{Akmal:1998cf} for the crust; but we also explore the impact of one
stiffer EoS model, ms1b \cite{Mueller:1996pm}, which
predicts a significantly ($2.6\times$) larger pressure at $\rho_{\rm sat}$ 
(see top panel of Fig.~\ref{fig:tabEOSs}).

The pressures at the
remaining five fiducial densities are free parameters, which we sample
uniformly via a Markov-chain Monte Carlo (MCMC). A tentative MCMC step is
rejected if it violates any of the minimal constraints enumerated in
Sec.~\ref{sec:intro_dopps}. We additionally impose a regularizer
on the pressure in our MCMC sampling,
in order to penalize EoS models that have extreme density variations in the
pressure.
This helps to compensate for the large degree of freedom
inherent to a five-polytrope EoS. We construct the regularizer to be a
Gaussian over the second logarithmic derivative of the pressure, i.e.,
\begin{equation}
\label{eq:reg}
    \xi = \exp \left[ - \frac{( d^2 (\ln{P})/d (\ln{\rho})^2 )^2} {2 \lambda^2} \right],
\end{equation}
where $\lambda$ is the characteristic scale. The distribution of $d^2
(\ln{P})/d (\ln{\rho})^2$ evaluated at different densities for a large
sample of theoretical EoSs is shown in Fig.~1 of \cite{Raithel:2017ity}.  Based
on those results, we adopt a conservative value of $\lambda=8$
for our baseline models, which corresponds to a weakly informative prior. We also
construct one set of models with $\lambda=2$, which corresponds
to taking stronger (more informative) input from the existing set of
nuclear models. For further details about how the choice of this Gaussian
regularizer affects an EoS inference, see \cite{Raithel:2017ity}.

Altogether, we construct five different sets of PWP EoSs:
our baseline model starts at $\rns$, uses ap3 for the crust EoS,
and adopts a weak prior on the second derivative of the pressures.
We explore two samples with the same crust and prior,
but starting the PWP parametrization at $\rho_0$= 1.2 and 1.5$\rns$.
We additionally modify the baseline sample to use ms1b for the crust EoS, 
keeping $\rho_0$ and the weak prior the same;
and to use a stronger prior, keeping $\rho_0$ and the crust EoS the same.
For each case, we generate a large number
(2-4 million) of parametric EoSs, in order to densely sample the 
five-dimensional EoS parameter space. We
summarize these samples in Table~\ref{table:cases} for reference in the following analysis.

    \begin{table}[!th]
    \centering
    \begin{tabularx}{0.45\textwidth}{@{\extracolsep{\fill}}cccc}
    \hline \hline
      Starting density & Crust EoS & Prior choice  \\ 
    \hline   
     $\rns$  & ap3 & Weak ($\lambda=8)$  \\ 
      {$1.2\rns$} &  {ap3} &  {Weak ($\lambda=8)$}   \\ 
     {$1.5\rns$} &  {ap3} &  {Weak ($\lambda=8)$}   \\
    \hline
      {$\rns$} &   {ms1b} &  {Weak ($\lambda=8)$}    \\ 
     {$\rns$} &  {ap3} &   {Strong ($\lambda=2)$}    \\
    \hline
    \end{tabularx}
    \caption{Overview of EoS populations considered in this work.}
      \label{table:cases}
\end{table}

\subsection{D\"oppelganger scoring metric}
\label{sec:score}
Within each EoS sample described in Table~\ref{table:cases}, we search for models that show
minimal differences in their tidal deformability curves, in spite of large
differences in the EoS. In particular, we take a subset of
$\sim9\times10^5$ models and compare each of these to every other EoS in
the full sample. For each possible pair of EoSs, we
define a ``\dop score" according to 
\begin{equation}
\label{eq:score}
\mathcal{D} = \exp\left[-\frac{(\Delta \Lambda_{\rm max})^2}{2\sigma_{\Lambda}^2}\right] \left\{1-\exp\left[-\frac{(\Delta R_{\rm min})^2}{2 \sigma_R^2}\right] \right\}
\end{equation}
where $\Delta \Lambda_{\rm max}$ is defined as the maximum difference
(i.e., the $L^{\infty}$-norm) in $\Lambda$ between the two EoSs  at any
mass across the astrophysically-observed mass range,
which we take to span from the lightest observed pulsar at
1.17~$\Ms$ \cite{Martinez:2015mya} to 2.01~$\Ms$
\cite{NANOGrav:2019jur,Fonseca:2021wxt}, at their 1-$\sigma$ lower limits.
We calculate $\Delta R_{\rm min}$ as the
minimum difference in radius between the two EoSs across the same mass
range. The first term in eq.~(\ref{eq:score}) is one for EoSs that predict
identical tidal deformabilities at all masses, and goes to zero as the
tidal deformability curves become less similar; while the second term is
zero for EoSs with identical mass-radius curves, and approaches one as the
mass-radius curves become more distinct. Thus, the \dop scoring criteria is
largest when $\Delta \Lambda_{\rm max}$ is \textit{minimized} and $\Delta
R_{\rm min}$ is \textit{maximized}. 
Finally, to set the scale in eq.~(\ref{eq:score}), we
define $\sigma_{\Lambda}=10$ and $\sigma_{R}=0.3$~km, based on the
characteristic differences that we found in $\Lambda$ and $R$ in
Sec.~\ref{sec:intro_dopps}.

We note that this scoring metric necessarily selects for \dops that are
similar to what we constructed by hand in Sec.~\ref{sec:intro_dopps}.
Other scoring criteria would select for different features. For example,
maximizing a mass-averaged $\Delta R$ (instead of $\Delta R_{\rm min}$)
tends to select for EoSs that are dominated by large differences in radius
near the maximum mass turn-over, even if the mass-radius curves are
identical at other masses.  It may also be possible to search for \dop
behavior directly in pressure-density space.  Empirically, we find that the
scoring criteria in eq.~(\ref{eq:score}) works well for identifying \dops
with similar phenomenological features to the examples shown in
Sec.~\ref{sec:intro_dopps}.

For each EoS in our $\sim 9\times10^5$ subset, we use eq.~(\ref{eq:score})
to identify a ``best" \dop companion out of all possible
(2-4 million) companions. For many EoSs, there is no high-scoring
\dop companion, and the ``best" score is correspondingly very low. However,
for the highest-scoring pairs of EoS, we find \dops that are comparable to
those constructed by hand in Figs.~\ref{fig:tabEOSs} and \ref{fig:degeneracy}.

\begin{figure}%[!ht]
\centering
\includegraphics[width=0.5\textwidth]{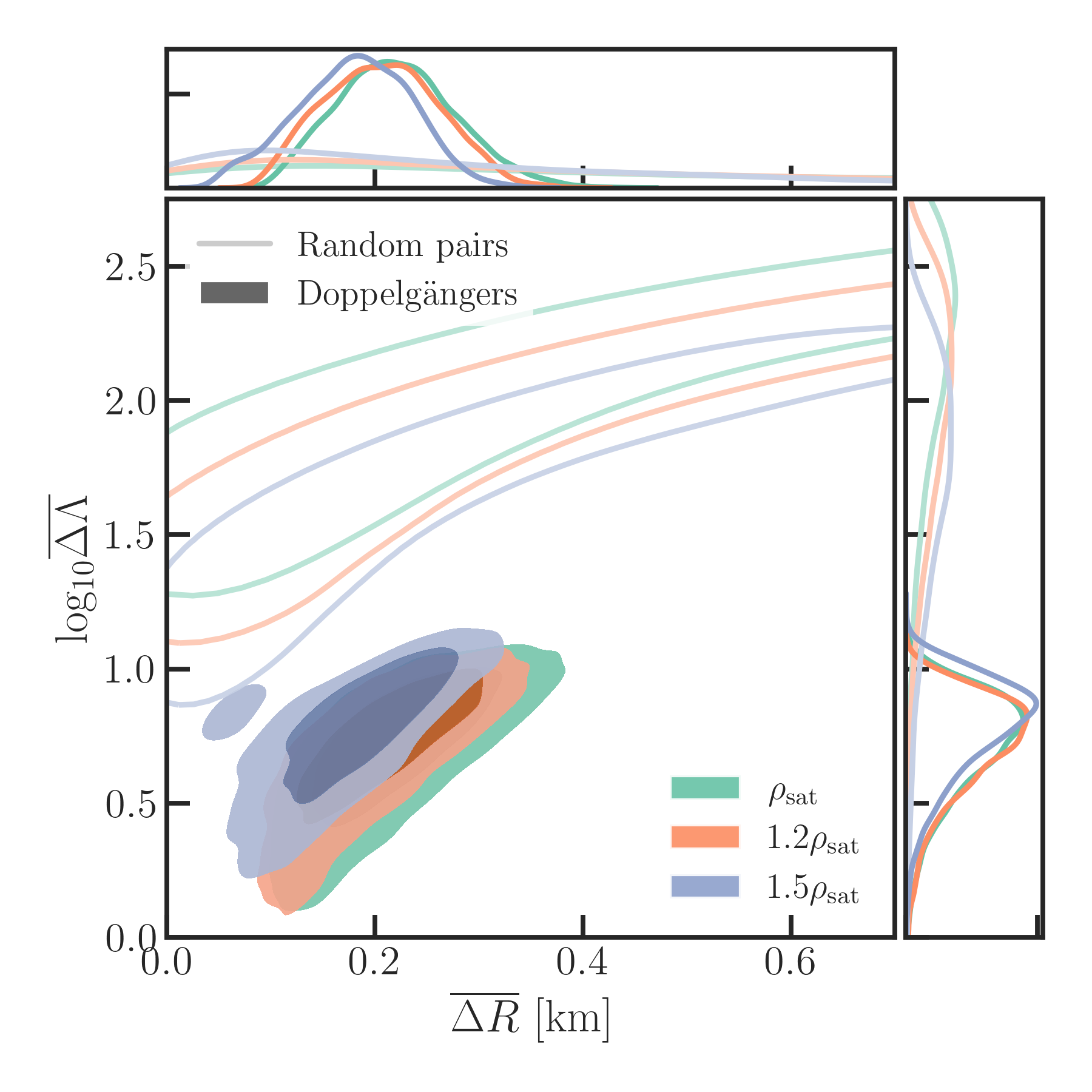}\\
\caption{\label{fig:dRdL} 
Contours showing the mass-averaged differences in radius
and tidal deformability, calculated across the mass range of
1.17-2.01$\Ms$. The filled contours show the
average differences for the highest-scoring set of \dop EoSs (68\% and 95\% intervals),
while the unfilled contours represent the average differences
for pairs of randomly-selected EoSs from each sample (68\% intervals).
We include results for three EoS samples, for which the
PWP parametrization starts at $\rho_0 = \{1,1.2,\text{and }1.5\}\rns$,
where $\rns$ is the nuclear saturation density.
All samples use the same crust EoS (ap3) at lower densities and
adopt the same weak ($\lambda=8$) prior on the pressure derivatives
at higher densities. As the crust EoS is assumed to higher densities,
the allowed \dops become less extreme.}
\end{figure}

\subsection{Population properties} 
\label{sec:population}
\begin{figure}%[!ht]
\centering
\includegraphics[width=0.45\textwidth]{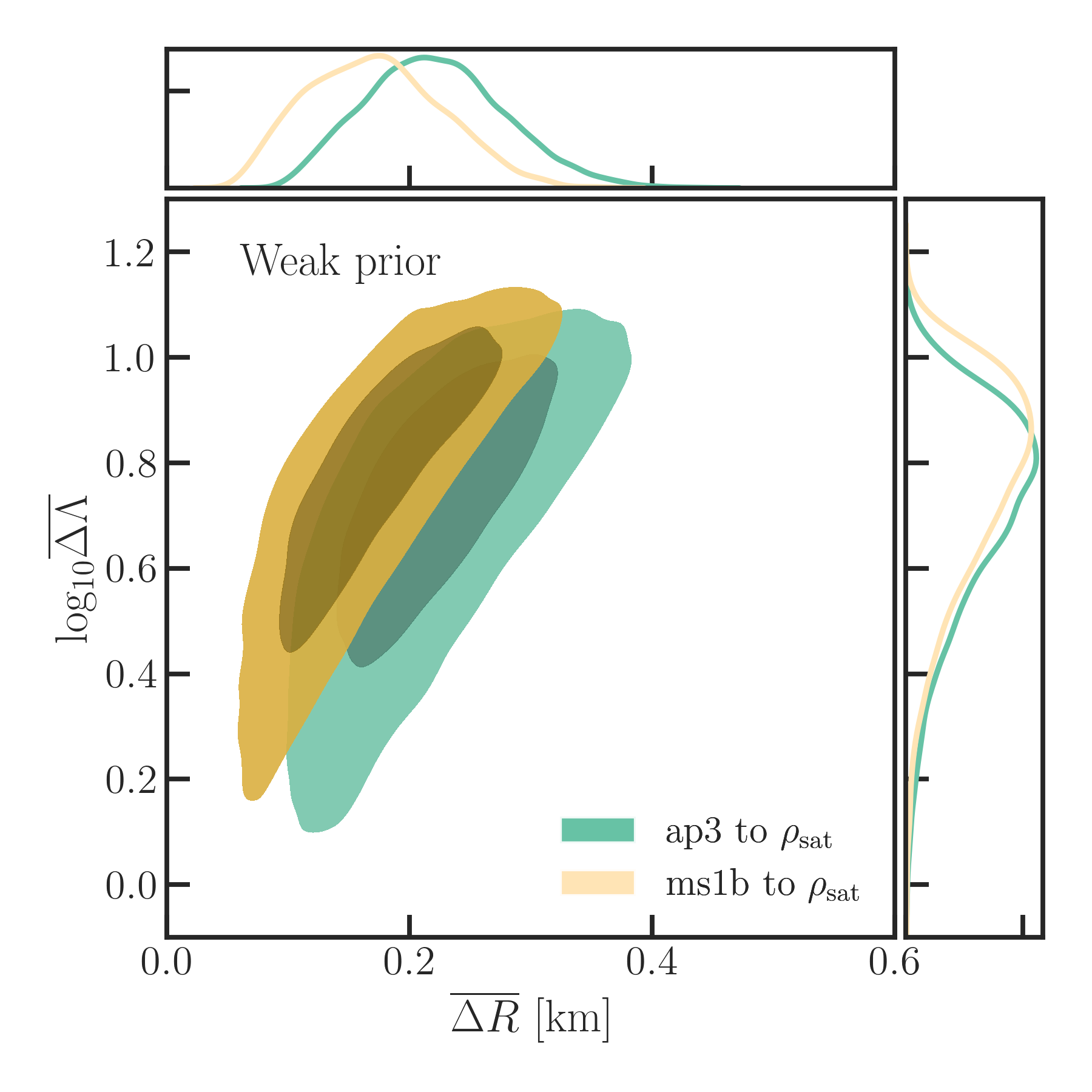}
\caption{\label{fig:dRdL_crust} Same as Fig.~\ref{fig:dRdL}, but showing the impact of the crust EoS on
the \dop populations. Both samples have a fiducial density of $\rho_0=\rns$ and a weak prior on
the second derivatives of the pressure. The baseline case using ap3 for the crust EoS is
repeated from Fig.~\ref{fig:dRdL} in teal; the sample using ms1b for the crust is shown
in yellow.}
\end{figure}

\begin{figure}%[!ht]
\centering
\includegraphics[width=0.45\textwidth]{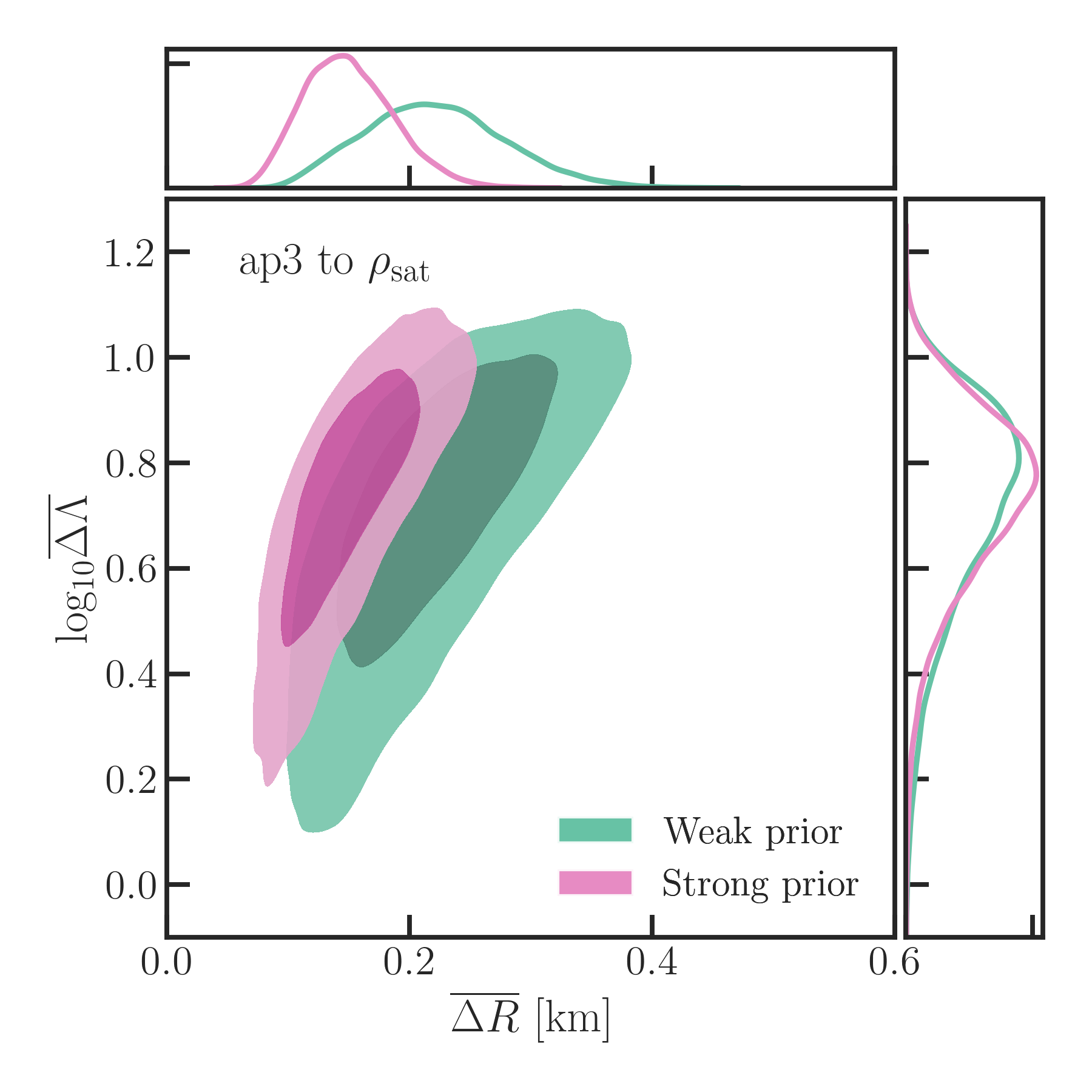}
\caption{\label{fig:dRdL_priors} Same as Fig.~\ref{fig:dRdL}, but showing the impact
of the choice of prior on
the \dop populations. Both samples have a fiducial density of $\rho_0=\rns$ and assume
ap3 for the crust EoS. The baseline case for a weak prior ($\lambda=8$) is
repeated from Fig.~\ref{fig:dRdL} in teal; the sample with a stronger prior $(\lambda=2$) is shown
in pink.}  
\end{figure}

We compute the distribution of scores $\mathcal{D}$
for each sample of randomly-generated EoSs
and we classify as \dops those pairs of EoSs that have  $\mathcal{D}$ within
90\% of the highest score found.
This criterion selects $\mathcal{O}(10^3 - 10^4)$ \dops for each sample of EoSs.
We show contours of the mass-averaged $\Delta \Lambda$ and $\Delta R$ for 
these models in Figs.~\ref{fig:dRdL}-\ref{fig:dRdL_priors}.  

To start, Fig.~\ref{fig:dRdL} shows the results for
the EoS samples that start their PWP parametrizations
at $\rho_0=\{1,1.2,\text{and } 1.5\}\rns$.
All three of these samples use the same low-density crust EoS (ap3) and the same weak prior
(see Table~\ref{table:cases}).
 For comparison, Fig.~\ref{fig:dRdL} also includes
 contours of the mass-averaged $\Delta
\Lambda$ and $\Delta R$ for 5,000 randomly-drawn pairs of
EoSs from each sample in the unfilled contours.

In general, randomly-paired EoSs have large differences in tidal
deformability, which are correlated with the difference in
radius, as one might typically expect.  In contrast, the samples of EoSs that
we identify as \dops occupy a distinct part of the EoS parameter space.
These have average differences in $\Lambda$ of $\lesssim10$,
indicating that the mass-tidal deformability curves of these pairs of EoSs
are extremely similar to one another, while the average difference in radius is
$\sim0.1-0.4$~km.
From Fig.~\ref{fig:dRdL}, we already see that the EoS
sample constructed with the PWP
parametrization starting at $\rns$ enables the most extreme set of
\dops, with average radius differences of up to $\sim$400~m. As
the crust EoS is enforced to higher densities, the allowed \dops become less extreme.

Figure~\ref{fig:dRdL_crust} shows the impact of assuming a different crust EoS
on the resulting sample of \dops. In this figure, the baseline EoS sample,
which assumes ap3 to $\rho_0=\rns$, is repeated in teal for reference.
The new EoS sample (shown in yellow) is
otherwise identical (i.e., using $\rho_0=\rns$ and the weak prior), 
but the low-density EoS
is exchanged for the stiffer model ms1b. 
We find that adopting a stiffer crust EoS results 
in less extreme \dops, characterized by smaller average differences
in radii.

Figure~\ref{fig:dRdL_priors} likewise shows
the impact of modifying the baseline EoS sample (in teal)
to use a stronger prior on the second logarithmic derivative of the 
pressure (in pink), such that $\lambda=2$ in eq.~\ref{eq:reg}. By restricting
rapid variations in the pressure as a function of density, the
extremity of the \dop models can be further reduced.

\begin{figure*}%[!ht]
\centering
\includegraphics[width=0.45\textwidth]{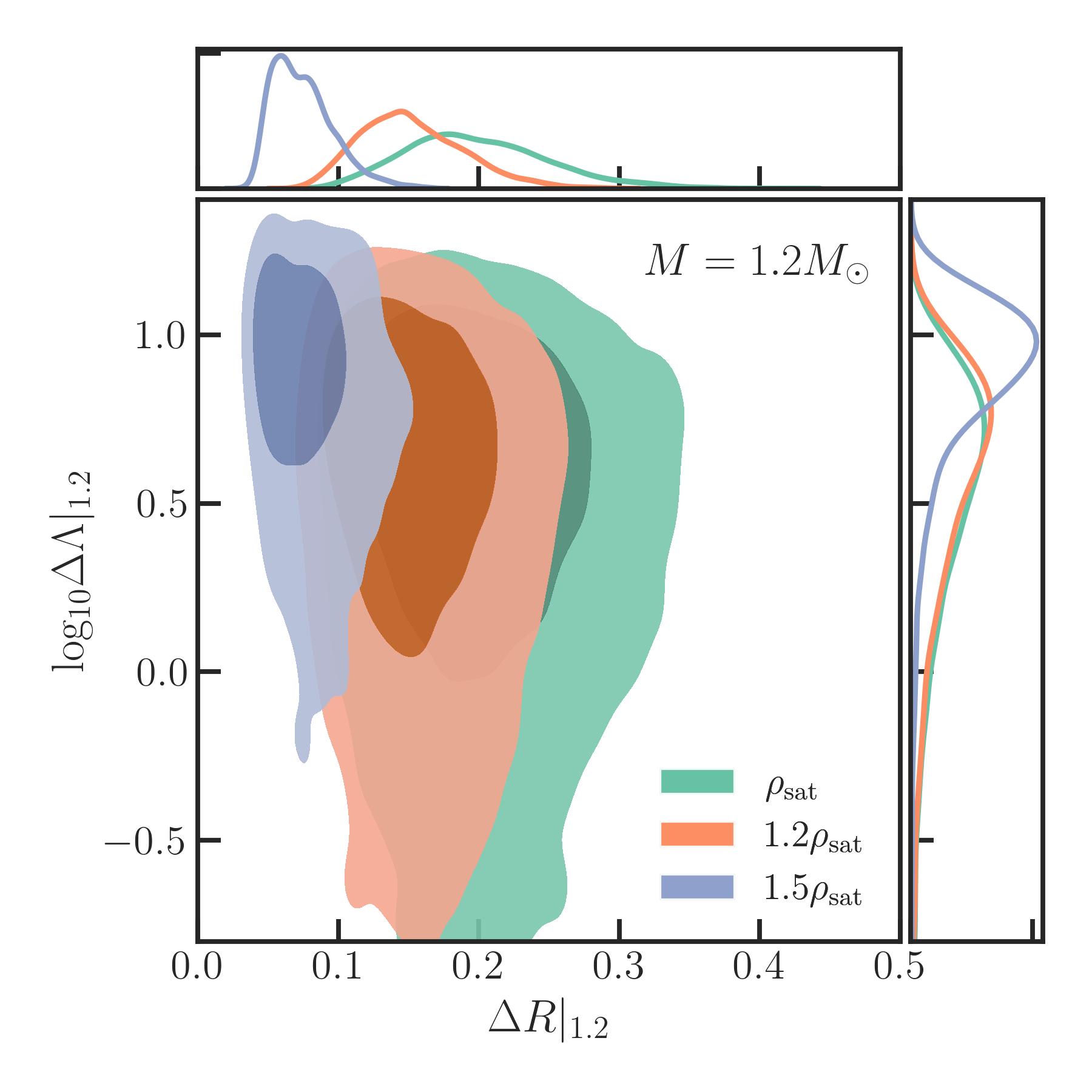}
\includegraphics[width=0.45\textwidth]{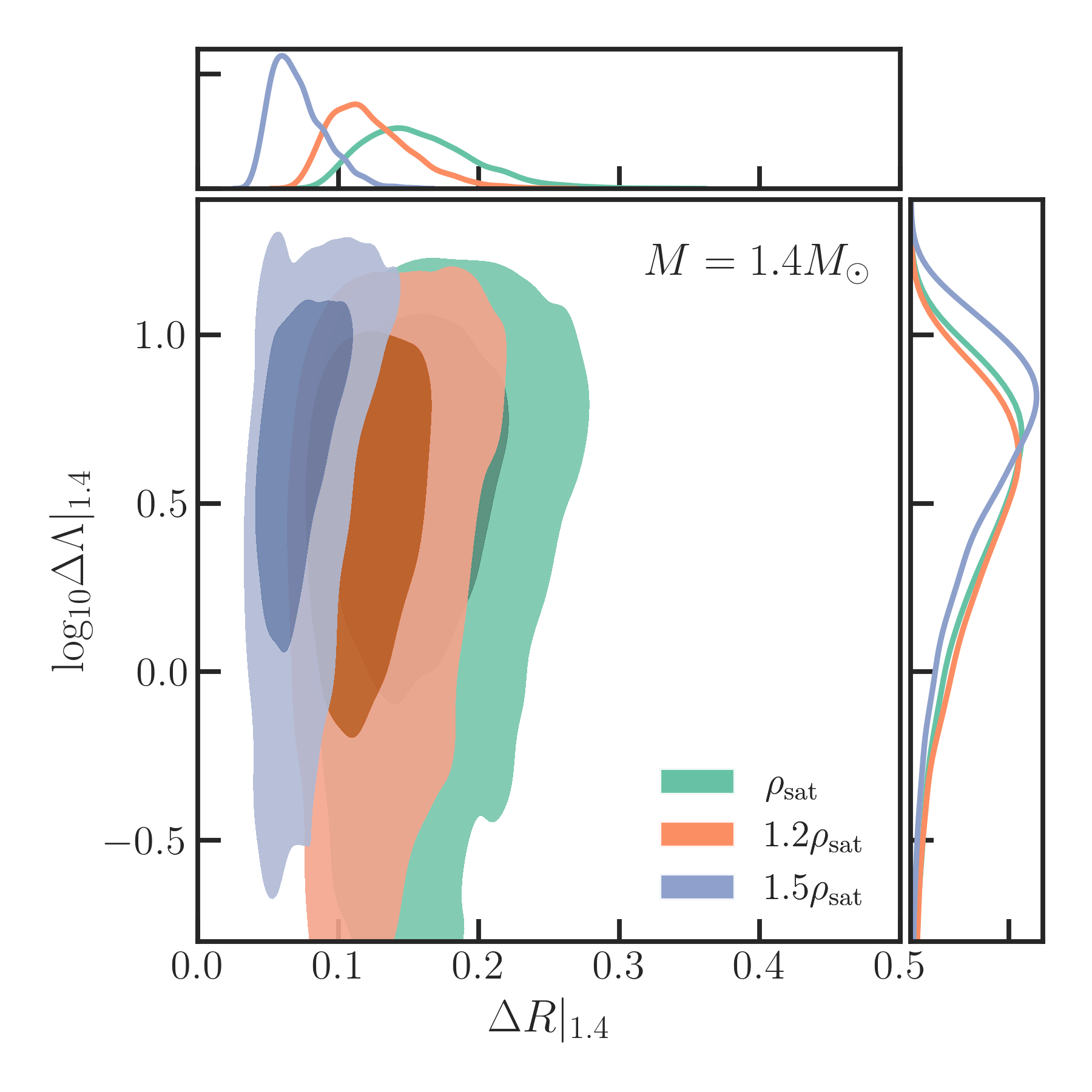}\\
\includegraphics[width=0.45\textwidth]{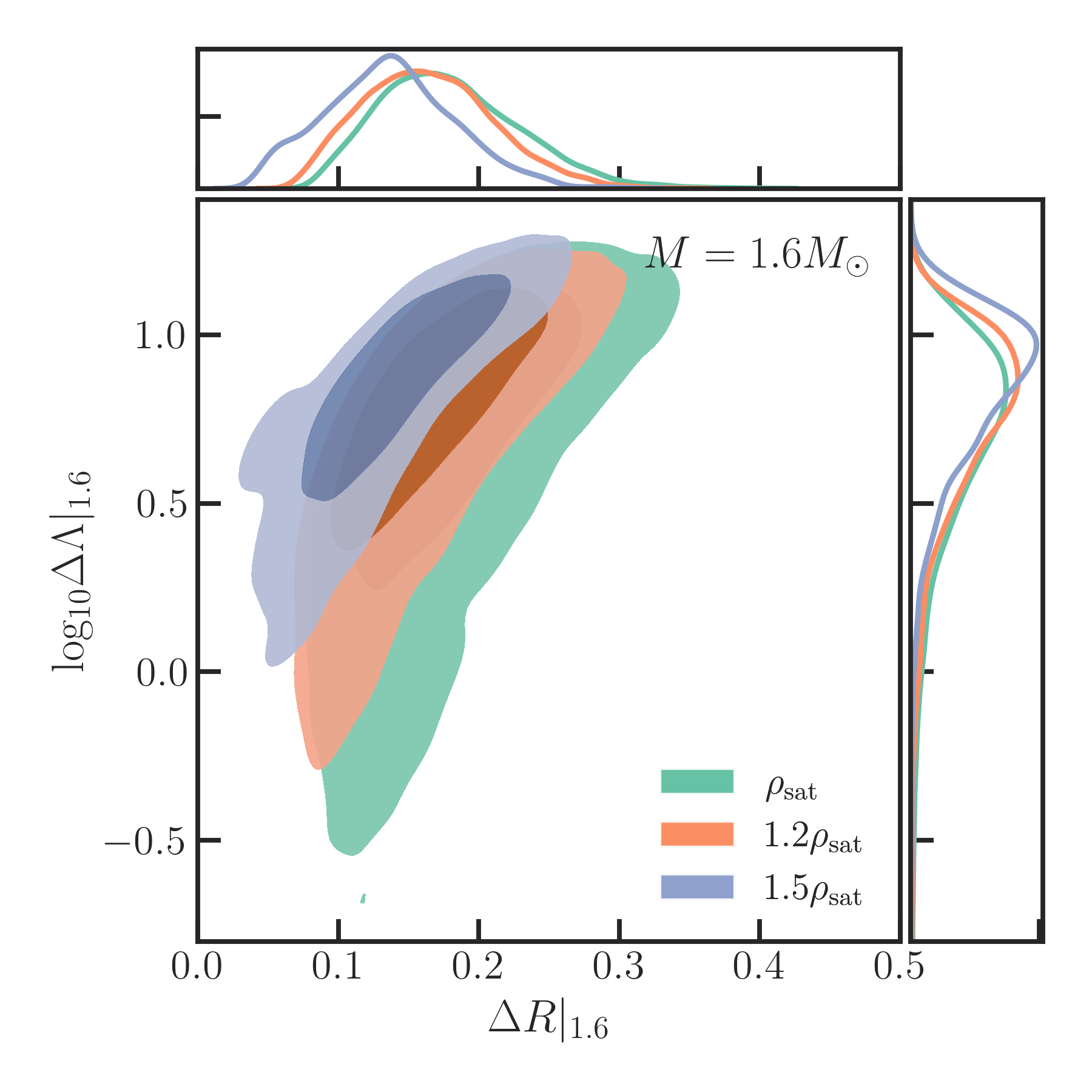}
\includegraphics[width=0.45\textwidth]{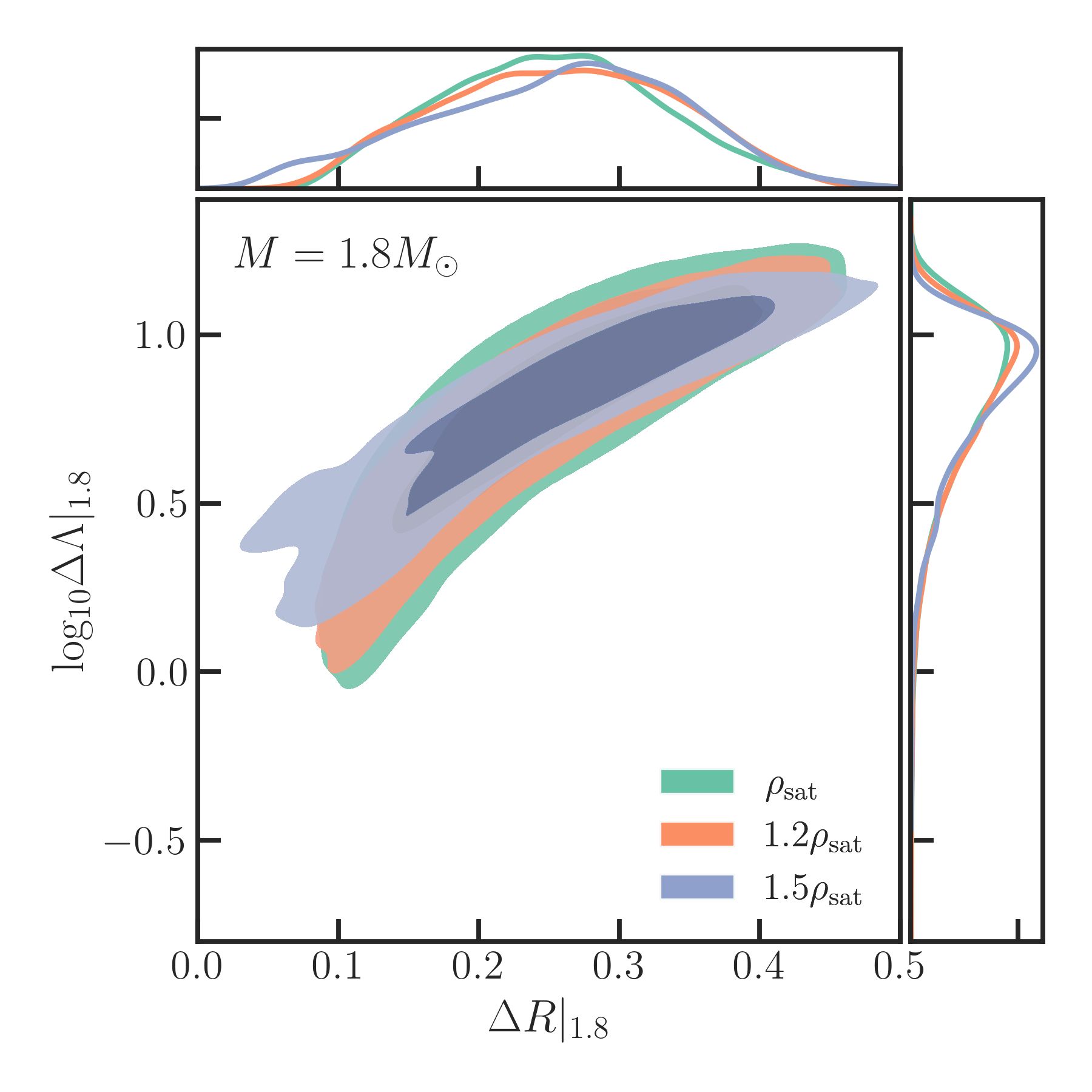}
\caption{\label{fig:dRdL_mass} 
Contours showing the absolute difference in $\Lambda$ and $R$ measured
at four different masses, for the \dops identified in Fig.~\ref{fig:dRdL}. The colors indicate the density at which the PWP parametrization begins, below which the crust EoS (ap3) is assumed. All models are sampled with the weak prior. The contours correspond to 68\% and 95\% intervals. The absolute differences in radii tend to be largest at small masses; while the differences in $\Lambda$ tend to be largest at higher masses. In all cases, $\Delta \Lambda$ measured at any of these masses is $\lesssim 20$.}
\end{figure*}

To understand these differences as a function of mass,
Fig.~\ref{fig:dRdL_mass} shows contours of the differences
in $\Lambda$ and $R$ at fixed masses, for each set of \dops identified
in Fig.~\ref{fig:dRdL}.
In general, we find that the \dops
selected for by eq.~(\ref{eq:score}) have the largest differences in radius
at low masses ($1.2\Ms$). At all four masses considered, the \dops differ
in $\Lambda$ by a similar degree, albeit with somewhat larger
characteristic differences reached at 1.8$\Ms$. In fact, we can see
that for this mass, $\Delta \Lambda$ is more strongly correlated
with $\Delta R$, whereas $\Delta \Lambda$ is almost independent of $\Delta R$ 
at lower masses. Figure~\ref{fig:dRdL_mass}
again highlights that assuming the crust EoS to higher densities
reduces the parameter space of the \dop EoSs. For example, when the
crust EoS is assumed only up to $\rns$, the \dops are
characterized by $\Delta
R_{1.4}\lesssim0.25$~km. However, in the more restrictive case that the EoS
is known to $1.5\rns$, the resulting set of
most extreme \dops have $\Delta R_{1.4}\lesssim0.1$~km.

In summary, for all EoS samples considered, we find
populations of \dop EoS models that have average differences in
the tidal deformability of $\lesssim10$, but that differ
in radii by up to a few hundred meters. The extremity of the
\dop behavior -- characterized by the average 
differences in radii -- can be reduced by enforcing
the crust EoS to higher densities, using a crust EoS
that is relatively stiff, or by adopting
stronger priors on the density-derivatives of the pressure.

\begin{figure}[!ht]
\centering
    \centering
    \includegraphics[width=0.45\textwidth]{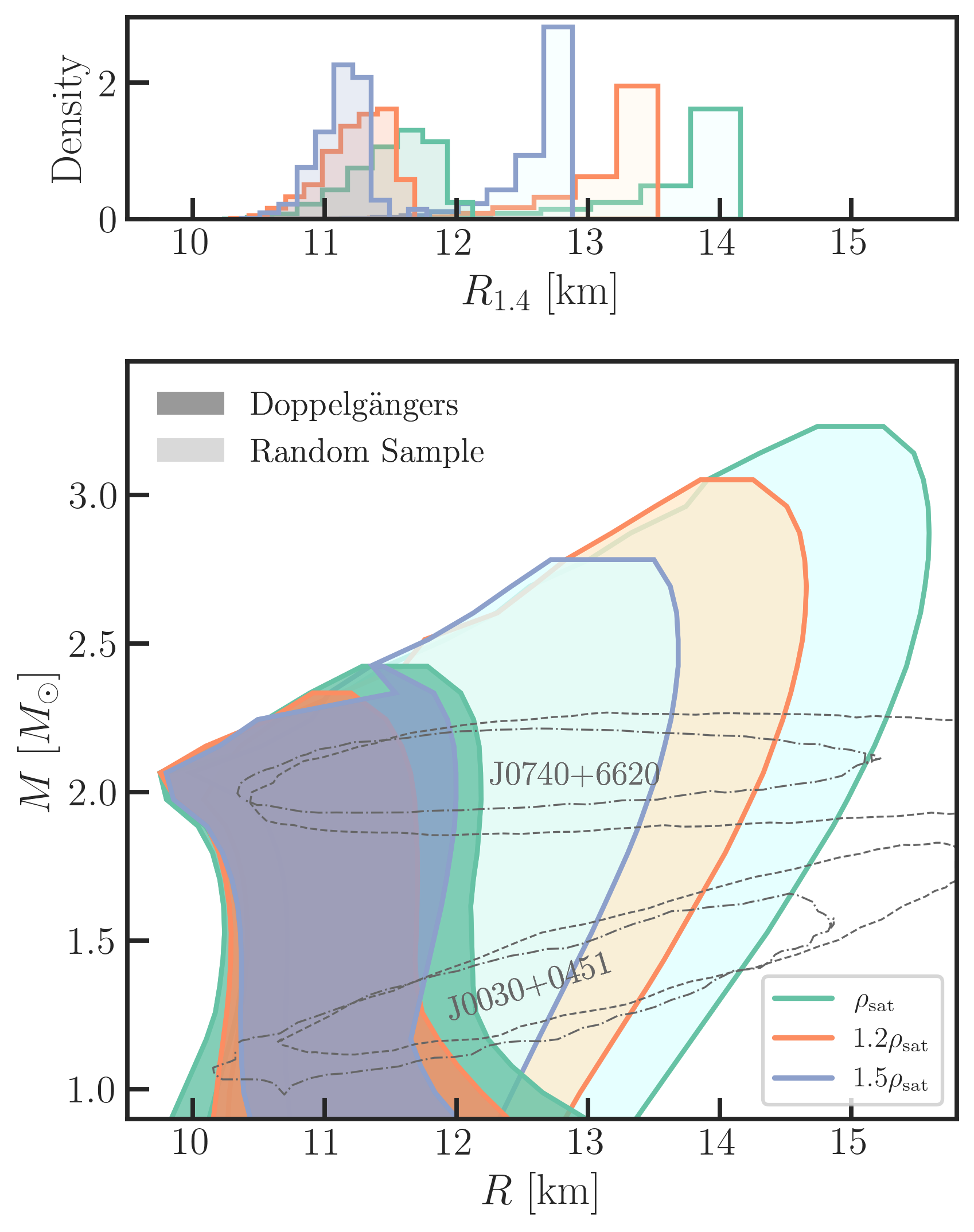}
    \caption{\label{fig:R14dist} Mass-radius space spanned by the EoS samples included in Fig.~\ref{fig:dRdL}. The EoS samples with the PWP parametrization starting from $\rns$ are shown in teal, from 1.2$\rns$ in orange, and from 1.5$\rns$ in blue. The light-shaded regions show a random sample of 5,000 EoSs, drawn from the complete distribution, while the dark-shaded regions correspond to the sample of highest-scoring \dop EoSs. 
    Mass-radius constraints from NICER are shown in gray dashed
    \cite{Miller:2019cac,Miller:2021qha} and dash-dotted lines \cite{Riley:2019yda,Riley:2021pdl}.
    The corresponding distribution of the radii predicted for a 1.4~$\Ms$ neutron star for each of these samples is shown in the top panel.
    The randomly-selected set of EoSs are strongly biased towards large radii, due to the uniform
    sampling in pressure-density space. In contrast, the \dop criterion selects for more compact stars.}
\end{figure}

\begin{figure}[!ht]
\centering
    \centering
    \includegraphics[width=0.45\textwidth]{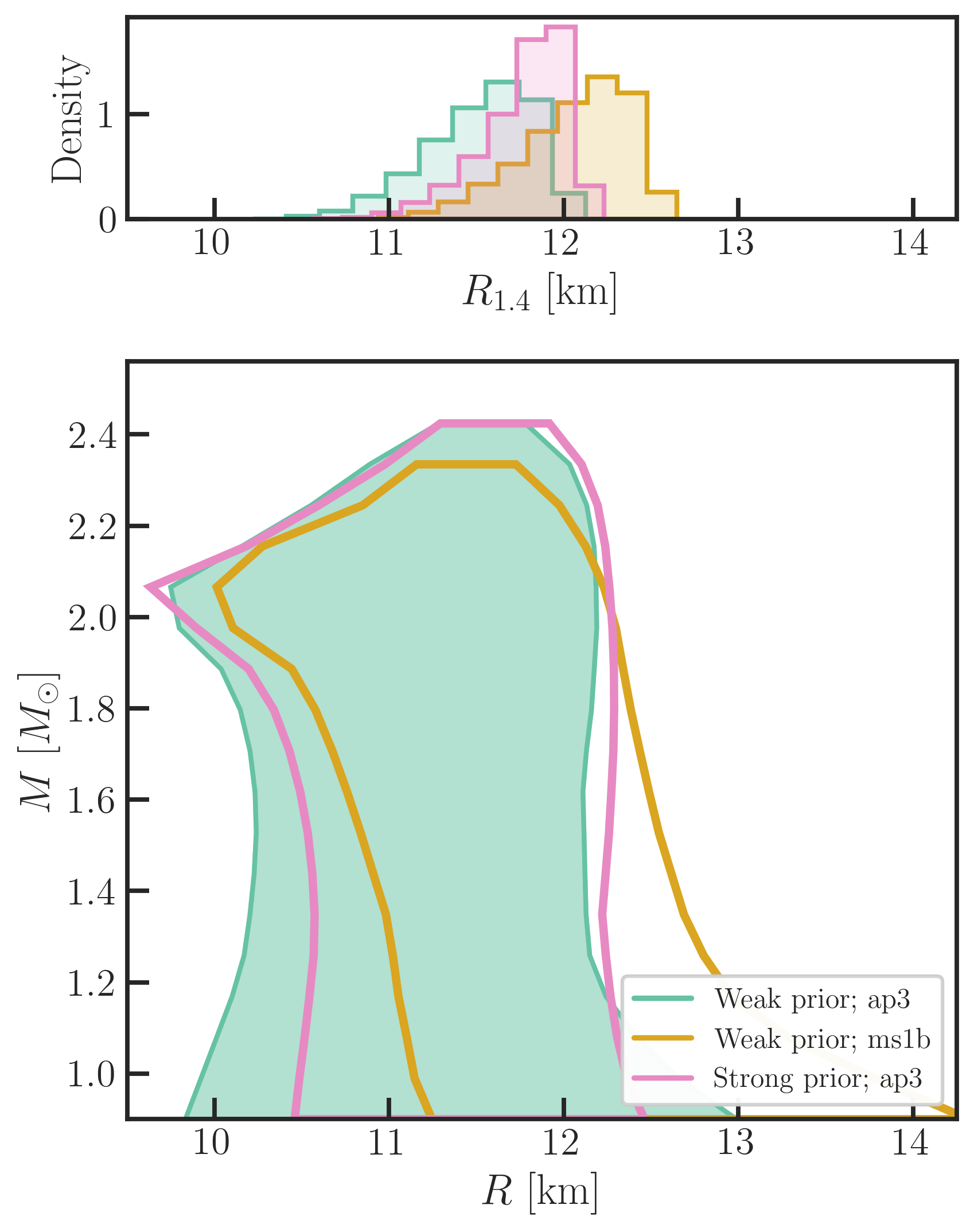}
    \caption{\label{fig:edges_priors} Same as Fig.~\ref{fig:R14dist}, but showing the impact of the choice of priors and the crust EoS on the \dop populations, compared to the baseline sample (in teal; repeated from Fig.~\ref{fig:R14dist}). All samples have a fiducial density of $\rho_0=\rns$. The impact of using the stiffer crust EoS ms1b is shown in yellow. The impact of adopting a stronger prior on the density-dependence of the pressure is shown in pink.} 
\end{figure}

\subsection{EoS parameter space of the \dop models}
\label{sec:space}

We turn now to the EoS parameter space probed by the tidal deformability
\dops. Figure~\ref{fig:R14dist} illustrates the mass-radius bounds spanned by each 
population of models from Fig.~\ref{fig:dRdL}.
The lighter shaded regions correspond to the
randomly-drawn EoSs from the full EoS samples, while the darker shaded
regions show the space occupied by the \dop EoSs.  We caution that these
are merely bounds on the mass-radius space, and that multiple different
EoSs may contribute to any given feature along these edges (see e.g. Fig.~3
of \cite{Raithel:2017ity}). However, these edges already demonstrate
clearly that the \dop scoring criterion selects for relatively compact
stars.

This is even more apparent when we consider the distributions of radii
for stars of fixed mass ($M=1.4\Ms$), which are
shown in the top panel of Fig.~\ref{fig:R14dist}.  The
randomly-drawn sample of EoSs peaks strongly towards large $R_{1.4}$, due
to the uniform MCMC sampling of the EoSs in pressure space. In spite of
this strong preference for large-radius stars in the randomly-drawn EoS
sample, the \dops select for more compact neutron stars, with
$R_{1.4}\simeq11-12$~km. As the starting density for
the PWP parametrization is increased, the distribution of \dops 
shifts toward smaller radii and spans a narrower region of the
mass-radius plane.

Figure~\ref{fig:edges_priors} shows the impact on these mass-radius bounds
of changing the crust EoS (in yellow) or adopting stronger priors (in pink),
compared to the baseline sample of \dops (in teal). For the same starting PWP density
of $\rho_0=\rns$, using a stiffer crust EoS such as ms1b causes
the \dop population to shift to larger radii, with the distribution of $R_{1.4}$
peaking at $\sim12.25$~km, compared to $R_{1.4}\sim11.7$~km for the
baseline sample that uses the softer crust EoS ap3.
When the stronger prior is adopted, the mass-radius bounds
shrink and $R_{1.4}$ is again slightly larger than
the weak prior baseline case. The reason for this is that the stronger
prior penalizes phase transitions (see eq.~\ref{eq:reg}),
which generally lead to more compact stars.

For the range of crust EoSs, priors, and starting densities considered here,
we find that the \dops occur in a relatively compact region
of parameter space. The range of stellar
compactness spanned by these models is
consistent with current astrophysical constraints 
from NICER \cite{Miller:2019cac,Riley:2019yda,Miller:2021qha,Riley:2021pdl}
and GW170187 \cite{LIGOScientific:2017vwq,LIGOScientific:2018hze,LIGOScientific:2018cki}.
However, if future observations find that neutron star radii
are significantly larger, it may be possible to constrain the ubiquity 
and extremity of the \dops by combining nuclear input (e.g., in terms of the densities to which
crust EoS is known) with X-ray observations of neutron stars. We revisit this point in the discussion below.

\begin{figure*}[!ht]
\centering
\includegraphics[width=\textwidth]{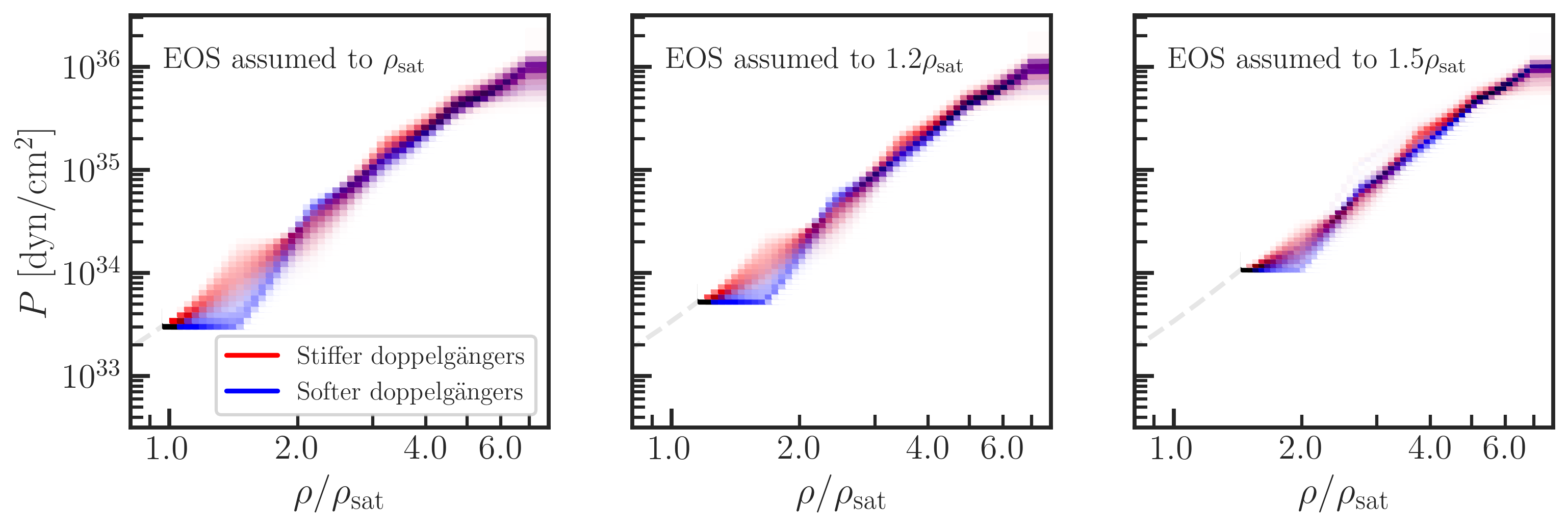}
\caption{\label{fig:p_dens} Pressure-density histograms for the population of \dops identified in Fig.~\ref{fig:dRdL}. All
models use ap3 for the crust EoS and use the weak prior on the density-dependence of the pressure.
We classify each EoS in a given pair of \dops as ``stiff" or ``soft" based on the 
pressure at the first fiducial density, and we plot the 2D histograms for each subclass in red and
blue respectively.}
\end{figure*}

\begin{figure}[!ht]
\centering
\includegraphics[width=0.425\textwidth]{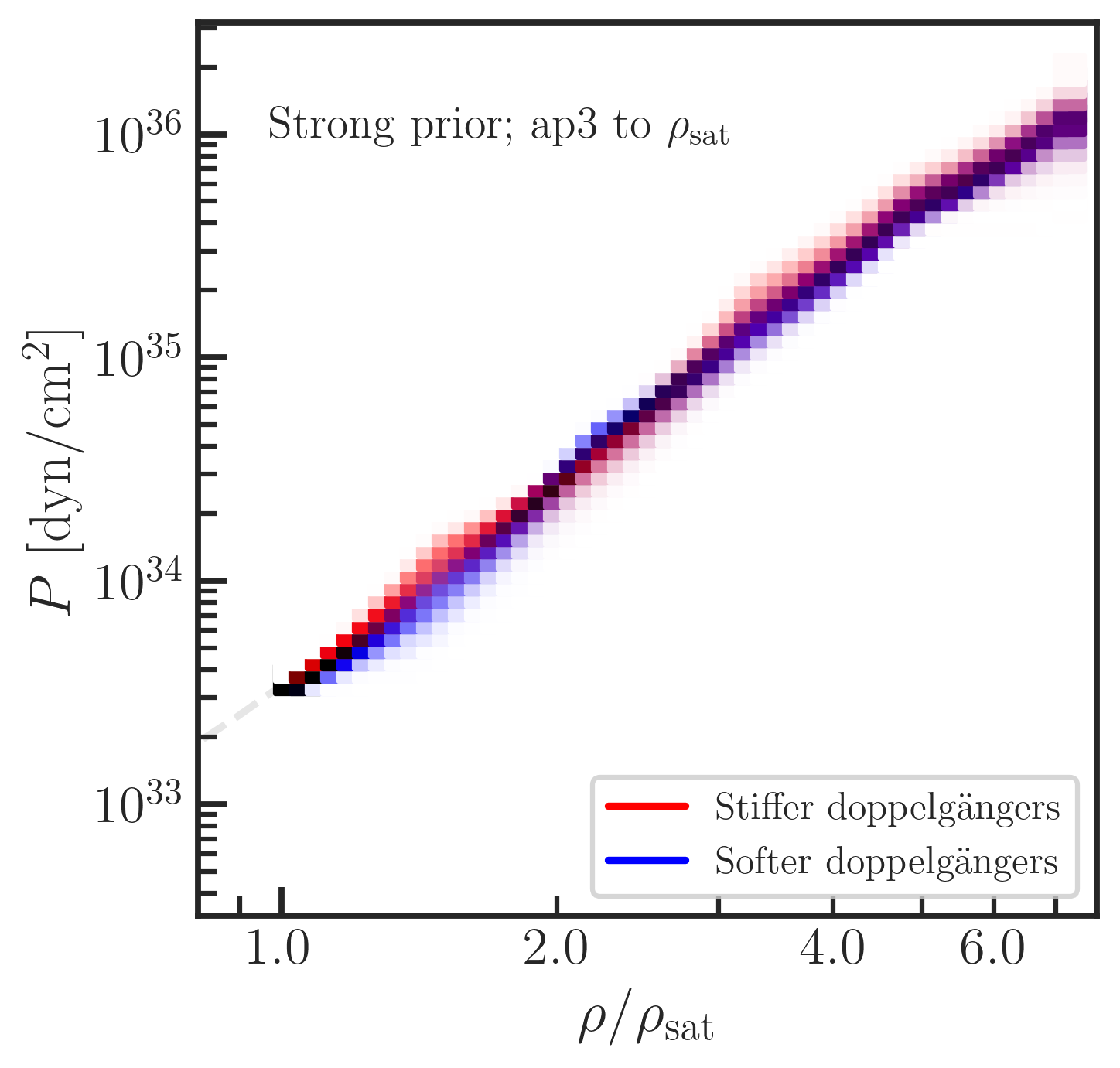}\\ 
\includegraphics[width=0.425\textwidth]{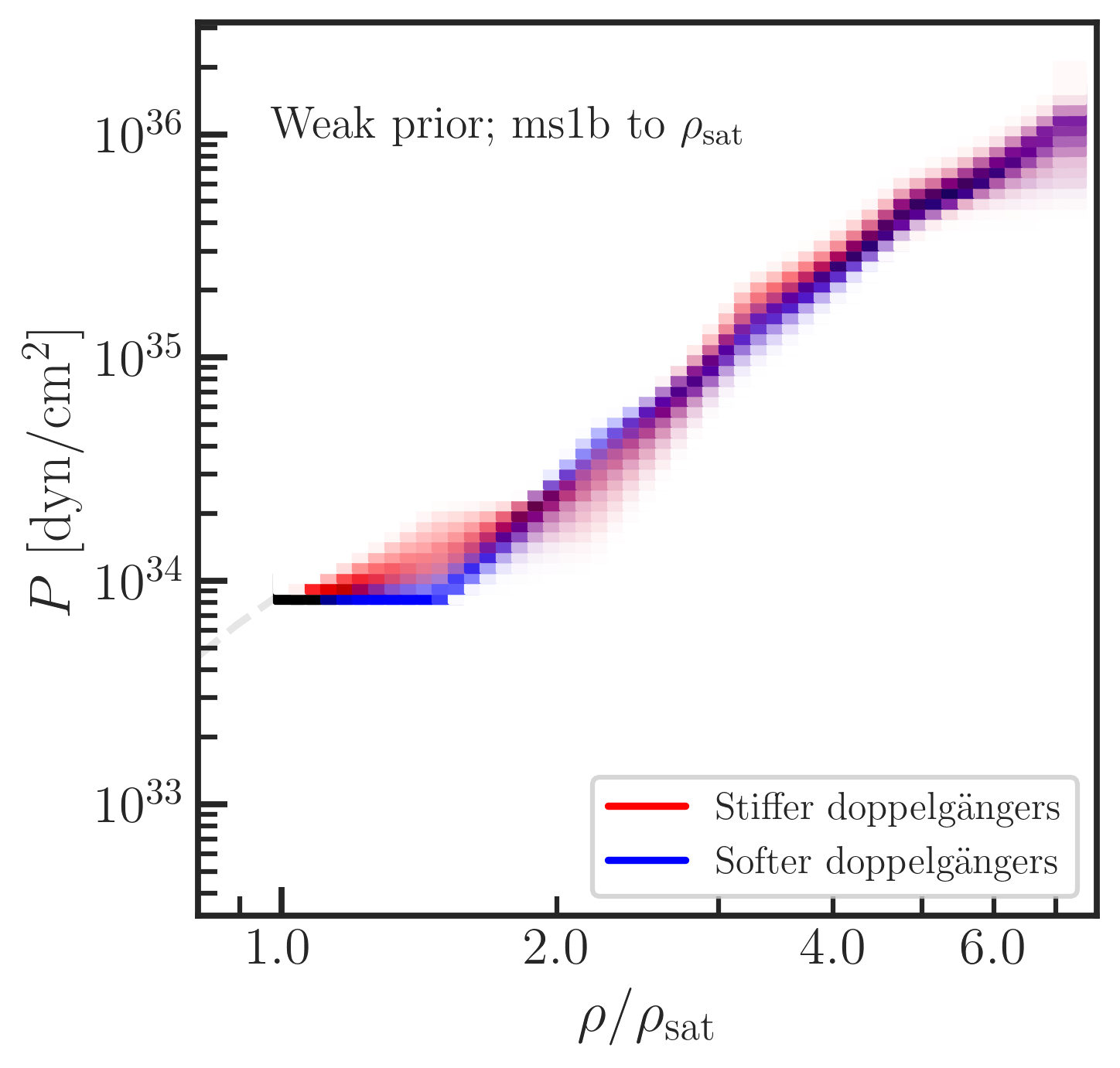} 
\caption{\label{fig:p_priors} 
 Same as Fig.~\ref{fig:p_dens}, but showing the impact of adopting a stronger prior (top) or stiffer crust EoS (bottom) on the population of highest-scoring \dops.
The PWP parametrization starts at $\rho_0=\rns$ for all models.}
\end{figure}

In order to better understand the parameter space of these models,
Figs.~\ref{fig:p_dens}-\ref{fig:p_priors} show
2D histograms of the pressure and density for the various populations of \dops. 
For a given pair of \dops, we identify one EoS
as ``softer" and one as ``stiffer", based on which has the larger pressure
at the first fiducial density in our parametrization. We then compute the 2D
histogram for all of the softer EoSs in blue, and the stiffer EoSs in red.
A clear structure emerges in these 2D histograms, with all softer (blue) EoSs
exhibiting some degree of a phase transition at densities immediately
above $\rho_0$, for any of the three starting densities considered in Fig.~\ref{fig:p_dens}. 
The stiffer (red) EoSs start initially stiff, then soften.
We saw a similar
behavior in Figs.~\ref{fig:tabEOSs} and \ref{fig:degeneracy}, where the strongest tidal
deformability degeneracy occurred for EoSs that differed maximally in the
pressure at low densities, but were similar at higher densities. Here, we
see that this signature holds more generally in this larger and
randomly-generated EoS sample: that is, the \dop degeneracy emerges for
EoSs that differ significantly at low densities, with one stiffer EoS
rising rapidly in pressure, undergoing a phase transition, and
then stiffening again (more slowly) to higher densities.  The companion
\dop EoS is one that starts with a strong phase transition
just above $\rho_0$, and then stiffens, such that
it has the higher pressure just above the first softening of the other
EoS. 

The large differences between the stiff (red) and soft (blue)
\dops in Fig.~\ref{fig:p_dens} imply significant differences
in the underlying physics of the models.
In particular, although the PWP
construction is phenomenological in nature, the qualitative features
found in Fig.~\ref{fig:p_dens}
are similar to what is found in more realistic calculations of EoSs with 
first-order phase transitions to deconfined quark matter,
or more generally to the emergence of more exotic degrees of freedom
\cite[e.g.,][and references therein]{Kojo:2014rca,Baym:2017whm,Blaschke:2018mqw}.
Thus, the degeneracy of the tidal deformability curves
for these models suggests some limitation to how well we may be able
to resolve phase transitions in some regions of parameter space.
We investigate this question in the context of mock EoS inferences
from GW data for \dop models
in \cite{Raithel:2022efm}.

As the crust EoS is assumed to higher densities, we find 
in Fig.~\ref{fig:p_dens} that the
differences in pressures between a given pair of \dops are reduced. 
For comparison against the baseline case (with the PWP parametrization 
starting at $\rho_0$), we show also in Fig.~\ref{fig:p_priors}
the \dop populations with the stronger
prior on the pressure derivatives and with the stiffer crust EoS.
For the case of the strong prior, which penalizes large
second-derivatives of the pressure (eq.~\ref{eq:reg}),
first-order phase transitions are severely restricted. Nevertheless,
the \dop scoring criteria still selects for EoSs that are maximally
different at intermediate densities and selects, when possible,
for smoother, cross-over phase transitions. 

For the crust EoS ms1b in the bottom panel of Fig.~\ref{fig:p_dens},
we also find smaller differences in the pressures between the stiff and
soft \dops, compared to
the baseline case which used ap3 for the crust. The reason for this 
is that ms1b is a relatively stiff EoS (see Fig.~\ref{fig:tabEOSs}),
thus the parametrization starts
from a higher pressure at $\rho_0$. As a result, there
is a narrower region of parameter space in which
it is possible to construct a pair of models
with the requisite phase transitions to achieve the
\dop morphology.

 We emphasize that these pressure-density
histograms highlight the most extreme pairs of \dops in our full sample.
As was shown in Fig.~\ref{fig:degeneracy}, the parameter space between the
blue and red curves will also be partially degenerate in tidal
deformability. We note that a different
parametrization of the EoS (e.g., a constant-sound-speed
\cite{Annala:2019puf},
spectral \cite{Lindblom:2018rfr,Lindblom:2022mkr} or
non-parametric \cite{Landry:2018prl,Legred:2022pyp} representation) may influence the EoS features
identified here and the resulting tidal deformability degeneracy as well. Further
work will be needed to assess the ubiquity of the \dop degeneracy in other EoS frameworks.

\subsection{I-Love-C Quasi-Universal Relations}
\label{sec:ILoveC}
The typical differences of $\lesssim10$ in the tidal deformability found in Sec.~\ref{sec:population}
are much smaller than expected
from the quasi-universal relation between $\Lambda$ and the stellar
compactness, $C$, first introduced in \cite{Yagi:2013awa,Yagi:2016bkt}. This $\Lambda-C$
relation predicts, e.g., that a 0.4~km difference in radii for
$R_{1.4}\simeq12$~km stars should propagate to a difference in
$\Lambda_{1.4}$ of 85. More generally, it has been shown that $\Lambda
\propto R^{\alpha}$ for a large range of EoSs, where $\alpha\approx6$ \cite{De:2018uhw,Raithel:2018ncd,Zhao:2018nyf}, which would further
imply that small differences in $R$ should propagate to large differences
in $\Lambda$. Given these standard scaling relations, the small differences
in $\Lambda$ for the \dops are quite surprising.

 \begin{figure*}[!ht]
\centering
\includegraphics[width=\textwidth]{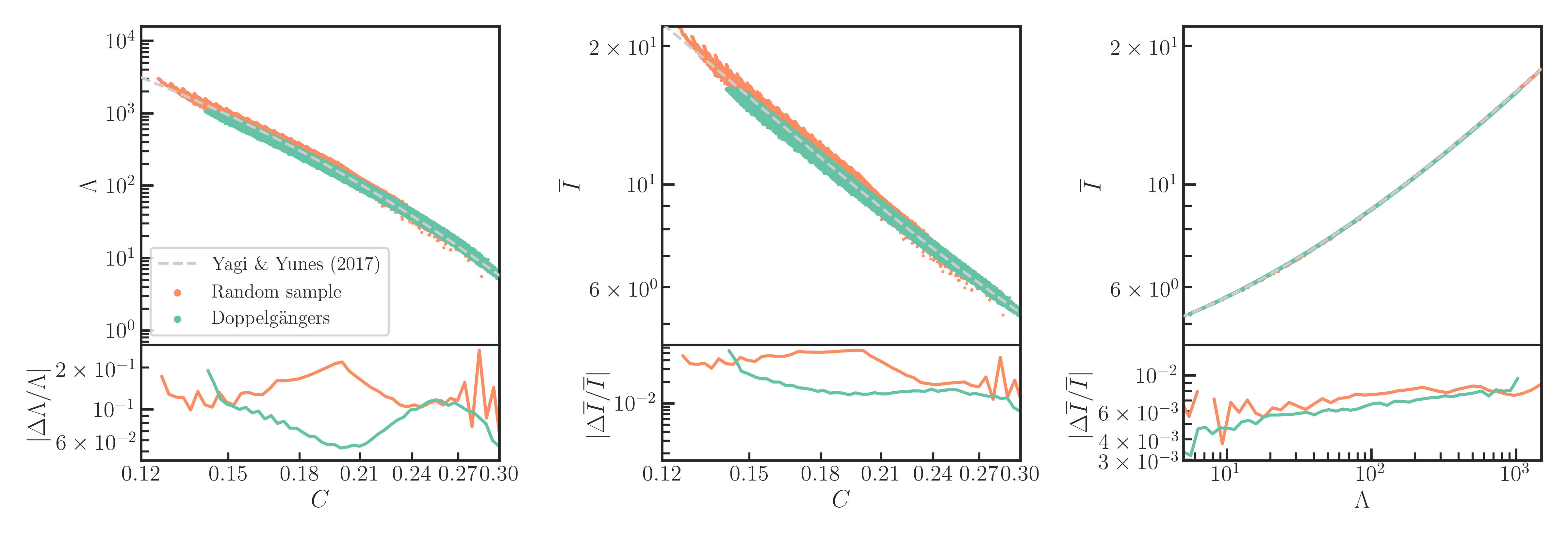}
\caption{\label{fig:ILoveC} Quasi-universal relation between tidal Love number, the neutron star compactness, $C$, and the dimensionless moment of inertia $\overline{I}\equiv I/M^3$.
A random selection of 5000 EoSs from our baseline EoS sample (which assumes the ap3 crust EoS to $\rns$ and a weak prior on the pressure derivatives) are shown in orange, while the set of most-extreme \dops are shown in teal. The quasi-universal relations from \cite{Yagi2017} are shown in gray, dashed lines for reference. The average residuals between the EoS samples and the quasi-universal fit line are shown in the bottom panels.}
\end{figure*}

Curiously, we find that the \dop EoSs still approximately obey the
quasi-universal relations of \cite{Yagi:2016bkt}, between not just
$\Lambda$ and $C$, but also with the dimensionless moment of inertia
$\bar{I}\equiv I/M^3$. We show these relations in Fig.~\ref{fig:ILoveC} for
our baseline EoS sample (PWP parametrization
starting at $\rns$, ap3 for the crust EoS, and a weak prior on the pressure derivatives).
The highest-scoring sample of \dops are shown in
teal, while the sample of randomly-drawn EoSs are shown in orange. The
bottom panel shows the residuals compared to the fit relations from
\cite{Yagi:2016bkt}. 

We find that the \dops obey these standard quasi-universal relations, with
a roughly comparable degree of scatter. In particular, for the
$\bar{I}-\Lambda$ relation, which is the tightest relation of the three
shown, the \dops obey the existing relation nearly exactly, with errors at
the sub-percent level. This is nearly indistinguishable from the relation
we find for the random sample of EoSs.  In the relations between either $\Lambda$
or $\bar{I}$ and the stellar
compactness, which are generally broader, we find that the \dops are approximately
consistent with the existing relationships, although there
is a systematic deviation from the standard relation at small compactness.

\begin{figure}[!ht]
\centering
\includegraphics[width=0.45\textwidth]{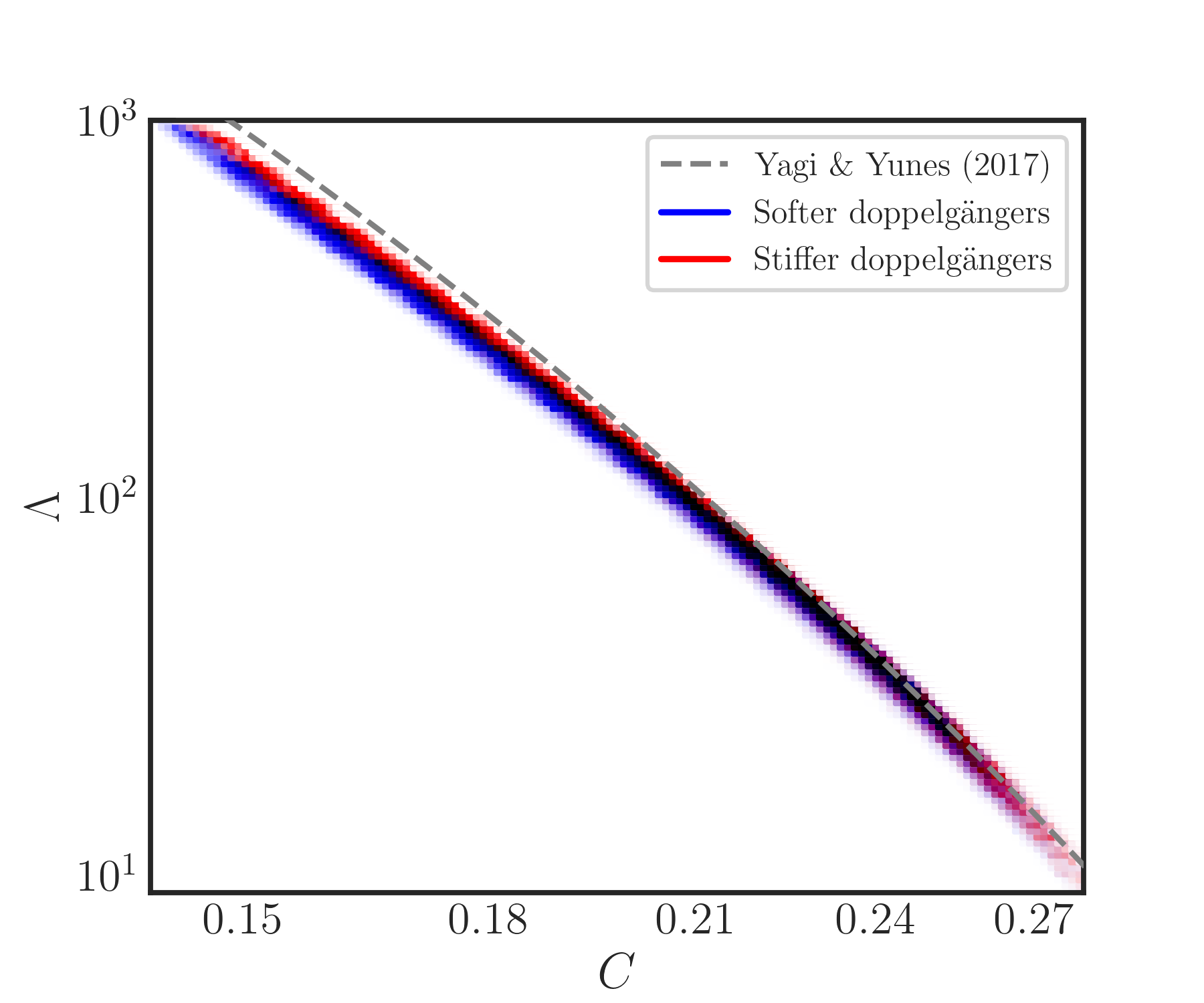}
\caption{\label{fig:loveC_hist} 2D histogram showing the quasi-universal relationship between
tidal deforambility and stellar compactness, for the \dops identified for our baseline EoS sample. We classify each EoS in a given pair of \dops as ``stiff" or ``soft" based on the 
pressure at the first fiducial density, and we plot the 2D histograms for each subclass in red and
blue respectively. Tidal deformability degeneracy emerges for EoSs that trace out parallel, but offset trends in $\Lambda-C$. }
\end{figure}

This deviation becomes more evident in Fig.~\ref{fig:loveC_hist},
where we show a 2D histogram of the $\Lambda-C$ relation for the same
sample of \dops.
As in Fig.~\ref{fig:p_dens}, for each pair of \dops, we
identify one softer and one stiffer EoS based on which has the larger
pressure at the first fiducial density and we color the resulting
histograms in blue and red, 
respectively. We find that the \dop models tend
to fall systematically below the $\Lambda-C$ relationship of \cite{Yagi:2016bkt},
with larger deviations at small $C$.
This is perhaps not surprising, given with the compact region of parameter space
that these models are found within (see Sec.~\ref{sec:space}).

We also find that, although the \dops generally obey the standard
$\Lambda-C$ relation, they select from distinct regions of this plane, 
as evidenced by the divide in colors in Fig.~\ref{fig:loveC_hist}.
That is, the \dops can be interpreted as originating from 
different sub-populations that exist within the scatter
of the broad $\Lambda-C$ relation. The results in Fig.~\ref{fig:loveC_hist}
are shown for the weakest set of constraints in order to illustrate the trend.

For completeness, we show the $I-\Lambda-C$ relations
for the EoS samples with additional restrictions (in terms of
the the density to which the crust EoS is assumed, the choice
of crust EoS, and the prior) in the Appendix. In all cases,
we find that the \dops exhibit a tight $I-\Lambda$ correlation, 
with sub-percent residuals comparable to the relation of \cite{Yagi:2016bkt}.
For the $\Lambda-C$ and $I-C$ relations, the \dop models
can deviate more significantly
from the relations of \cite{Yagi:2016bkt}, especially at low $C$,
depending on
the restrictions adopted in the EoS sampling.

\begin{figure}[!ht]
\centering
    \centering
    \includegraphics[width=0.45\textwidth]{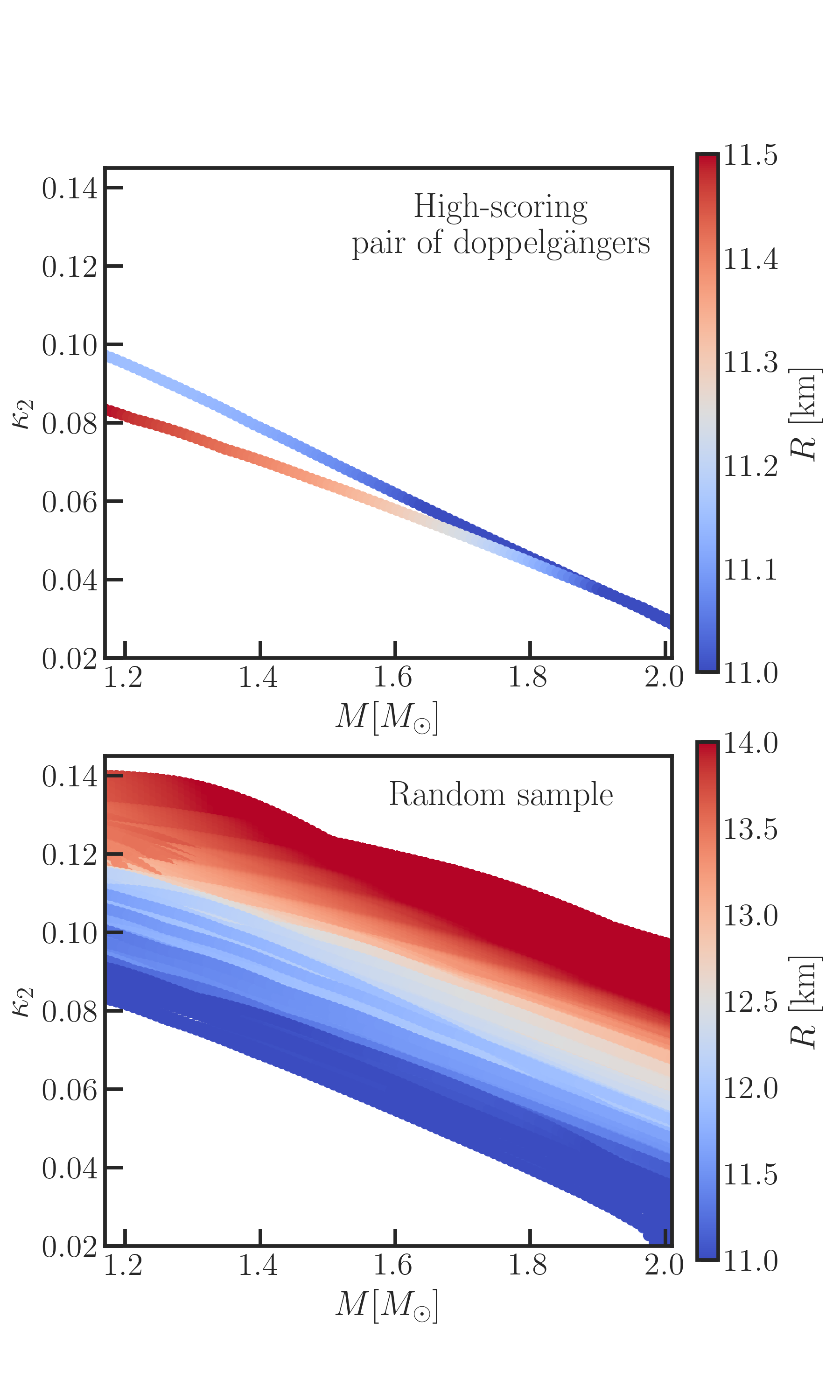}
    \caption{\label{fig:k2} Tidal Love number as a function of the neutron
  star mass and radius (shown via the color). The top panel shows these
values for one of the highest-scoring examples of \dops, while the bottom
panel shows the correlations for a randomly drawn sample of EoSs. All
results are shown for the baseline sample, which uses ap3 for the crust EoS up to
$\rns$, together with the weak prior on the pressure derivatives.}
\end{figure}

Finally, we illustrate the origin of the tidal deformability degeneracy in
Fig.~\ref{fig:k2}, where we show the tidal Love number, $k_2$, for one pair
of high-scoring \dops (in the top panel), compared to the random population
of EoSs (in the bottom panel). We see in this figure that, although the
tidal deformabilities of the two \dop EoSs are very similar, their tidal
Love numbers differ significantly. Figure~\ref{fig:k2} additionally shows
that for a random sample of EoSs, $k_2$ is approximately correlated with
the neutron star radius. Because the tidal deformability is constructed
from these two quantities according to 
\begin{equation}
\Lambda = \frac{2}{3} k_2 \left(\frac{R}{M}\right)^5,
\end{equation}
the bottom panel of Fig.~\ref{fig:k2} thus also illustrates the known
$\Lambda \propto R^{\alpha}$ scaling relation.

In contrast, the top panel of Fig.~\ref{fig:k2} shows that the \dops
violate this trend between $k_2$ and the radius, across a limited range of
radii. For the \dop models, an increase in $k_2$ at a given mass
is balanced by a
decrease in $R$, and the overall $\Lambda$ remains
approximately unchanged.  These results suggest that if a gravitational
wave detector were directly sensitive to $k_2$, the \dop models could be
more readily distinguished. However, because the detectors are most
sensitive to the product of $k_2$ and $R^5$, and because $k_2$ and $R$ are
each sensitive to the EoS at slightly different densities, it is possible
to construct models that are approximately degenerate in their tidal
deformabilities. We note that these trends hold generally for all of the EoS samples considered, but with differing magnitudes. We show results here for the baseline sample
for illustrative effect.

\section{Discussion}
\label{sec:discussion}

The example \dops first constructed by hand in Sec.~\ref{sec:intro_dopps} and
identified generically from samples of randomly-generated EoSs in Sec.~\ref{sec:param}
both share a common signature: namely, that the stiffer EoS undergoes some
cross-over (or, in the more extreme case, first-order) phase transition at 
supranuclear densities.
When compared to
a sufficiently soft companion EoS, the resulting models can have nearly identical
tidal deformability curves across a wide range of neutron star masses, in spite of
significant differences ($>100\%$) in the low-density pressures. We see
this behavior in EoS samples that are constructed to obey minimal physical constraints
and that are subject to different degrees of nuclear input, in terms of the density
to which the crust EoS is utilized and in terms of the prior on the derivatives of the pressure. 
In every case, we find examples of \dop EoSs, although the extremity of the degeneracy 
is reduced as more restrictions are added.

The existence of \dop EoSs thus seems to be a natural consequence of
allowing for a phase transition. For an EoS to undergo a
phase transition and still meet the maximum mass requirement of
$\sim2\Ms$, it requires a significant degree of stiffening at both lower
and higher densities. As we have shown in Fig.~\ref{fig:p_priors}, one way to
reduce the extremity of the allowed \dops is to apply a stronger prior on
the second derivative of the pressure. Thus, by
folding in constraints on the sound speed in neutron stars
from latest astrophysical measurements, combined
with theoretical input from chiral EFT 
 \cite[e.g.,][]{Tews:2018kmu,Reed:2019ezm,Drischler:2021bup,Altiparmak:2022bke},
 it may be possible to constrain the parameter space
of \dop models.

In terms of the crust EoS, further constraints may be possible
by incorporating latest results from chiral EFT
calculations, which can at least partially constrain the EoS to densities of 2$\rns$
\cite{Hebeler:2015hla,Lynn:2019rdt,Drischler:2021kxf},
Improvements in low-density experimental constraints will come in
the coming years as well, e.g., from measurements of the
neutron skin thickness with PREX and CREX \cite{Thiel:2019tkm}, new
neutron-rich isotope facilities such as FRIB, RIBF, and FAIR, and
next-generation heavy ion colliders such as NICA  \cite{Schatz:2022vzq}. As
these experiments and theoretical inputs constrain the crust EoS to higher
densities, the allowed parameter space for the \dop models will be further
reduced.

Another avenue forward is to use the next-generation of astrophysical
constraints. Indeed, the most extreme examples of \dops that we constructed
predicted neutron star radii that differ by $\sim0.5$~km (or even 0.7~km,
when allowing for further freedom in the crust EoS, as in
Sec.~\ref{sec:intro_dopps}). This is comparable to the anticipated radius
accuracy for the brightest NICER targets \cite{2012SPIE.8443E..13G}.
Gravitational wave detections of mergers with very low-mass neutron stars
($\sim1-1.2\Ms$) may also help to distinguish between some classes of \dops,
 as low-mass stars have generally larger differences
in tidal deformabilities. If such low-mass systems
exist, they may be a prime target for partially resolving the tidal
deformability degeneracy directly with GW data. We demonstrate this
prospect in the context of an EoS inference from current
and upcoming GW data for a pair of \dop models in \cite{Raithel:2022efm}.
However, we note that
measuring the difference in $\Lambda$ for such systems will likely still
require very high SNRs, at which point current waveform models become prone
to systematic errors \cite{Gamba:2020wgg}. Improvements in numerical waveform models are
thus crucially needed as well.

Rather than treating each of these types of constraints separately, the
most promising route forward will likely be to combine these inputs. For
example, consider a scenario in which nuclear theory constrains the crust
EoS sufficiently tightly that the allowed parameter space of \dops is
restricted to $R_{1.4}\in(10,12)$~km (as in e.g., Fig.~\ref{fig:R14dist}).
If NICER measures the radius of a future source with high precision to
$R_{1.4}>12$~km or if LIGO likewise measures an incompatible tidal
deformability, then the most extreme sets of \dops can effectively be ruled
out.

In a companion work to this paper \cite{Raithel:2022efm}, we performed a set of
numerical relativity simulations of neutron star mergers, using two example
pairs of \dop EoSs.  In that work, we investigated the emission of
gravitational waves from the post-merger remnant, and found that the peak
frequency of the post-merger  signal is the same for a given pair of \dops,
to within our estimate of the numerical error of the simulations. In other
words, it seems that post-merger GWs may not be able to
distinguish between \dops either, although this conclusion is subject to
the numerical and physical uncertainties of current state-of-the-art
simulations, including the possible impact of incorporating more realistic
magnetic fields \cite{Kiuchi:2015sga,Kiuchi:2017zzg,Palenzuela:2021gdo}, 
more advanced neutrino physics \cite{Alford:2017rxf,Most:2021zvc,Radice:2021jtw,Most:2022yhe},
finite-temperature effects \cite{Bauswein:2010dn,Figura:2020fkj,Raithel:2021hye}, and rapid neutron
star spins
\cite{Dietrich:2016lyp,East:2019lbk,Most:2019pac,Tsokaros:2020hli,Papenfort:2022ywx}, which will need to be considered before a final
conclusion can be drawn.

\section{Summary and Conclusions}

In this paper, we have introduced a new class of EoS models that can differ significantly in the pressure near saturation densities and, accordingly, in the radii by up to 0.5~km, but that are surprisingly similar in tidal deformability across the entire range of neutron star masses. These tidal deformability \dops will be challenging to differentiate with the current generation of GW detectors, although next-generation facilities such as Cosmic Explorer or Einstein Telescope may be able to resolve the small differences in $\Lambda$ for these models. 

We have shown that it is not only possible to construct EoSs that have nearly-degenerate tidal deformability curves, but that these \dops naturally occur in randomly-generated samples of EoS models, as a consequence of allowing for a phase transition at supranuclear densities, where the exact density can be pushed higher by adopting more restrictive nuclear input. This has important implications for EoS inferences from measurements of the tidal deformability. In particular, even with measurements of tidal deformabilities across a wide range of masses,
we have shown that there are some regions of the EoS parameter that will be challenging to distinguish based on 
the tidal deformabilities alone. This implies a fundamental limit to the level at which the neutron star radius can be constrained from current measurements of the tidal deformability, in the absence of informative nuclear priors that would distinguish between these \dop models.

However, we have also demonstrated that by adopting more restrictive priors on the density-dependence of the pressure, or by utilizing the crust EoS to higher densities, that the extremity of the \dop models can be significantly reduced. Thus, by incorporating future constraints from nuclear theory and experiments, X-ray observations of neutron star radii, and population constraints on the tidal deformability from current-generation facilities, it may be possible to significantly constrain the ubiquity of \dop models, even before the advent of next-generation GW detectors.

\begin{acknowledgments}
The authors thank
Gabriele Bozzola, Katerina Chatziioannou, Pierre Christian, Philippe Landry,
Feryal \"Ozel, Dimitrios Psaltis, Jocelyn Read, Ingo Tews, and Nicolas Yunes 
for insightful comments on this work.
The authors gratefully acknowledge support from postdoctoral fellowships at the Princeton Center for Theoretical Science, the Princeton Gravity Initiative, and the Institute for Advanced Study. 
CAR additionally acknowledges support as a John
N. Bahcall Fellow at the Institute for Advanced Study.
This work was performed in part at the Aspen Center for Physics, which is supported by National Science Foundation grant PHY-1607611.
The EoS parameter survey was performed on computational resources managed and supported by Princeton Research Computing, a consortium of groups including the Princeton Institute for Computational Science and Engineering (PICSciE) and the Office of Information Technology's High Performance Computing Center and Visualization Laboratory at Princeton University.
\end{acknowledgments}

\appendix

\section{I-Love-C relations for additional EoS samples}
\begin{figure*}[!ht]
\centering
\includegraphics[width=\textwidth]{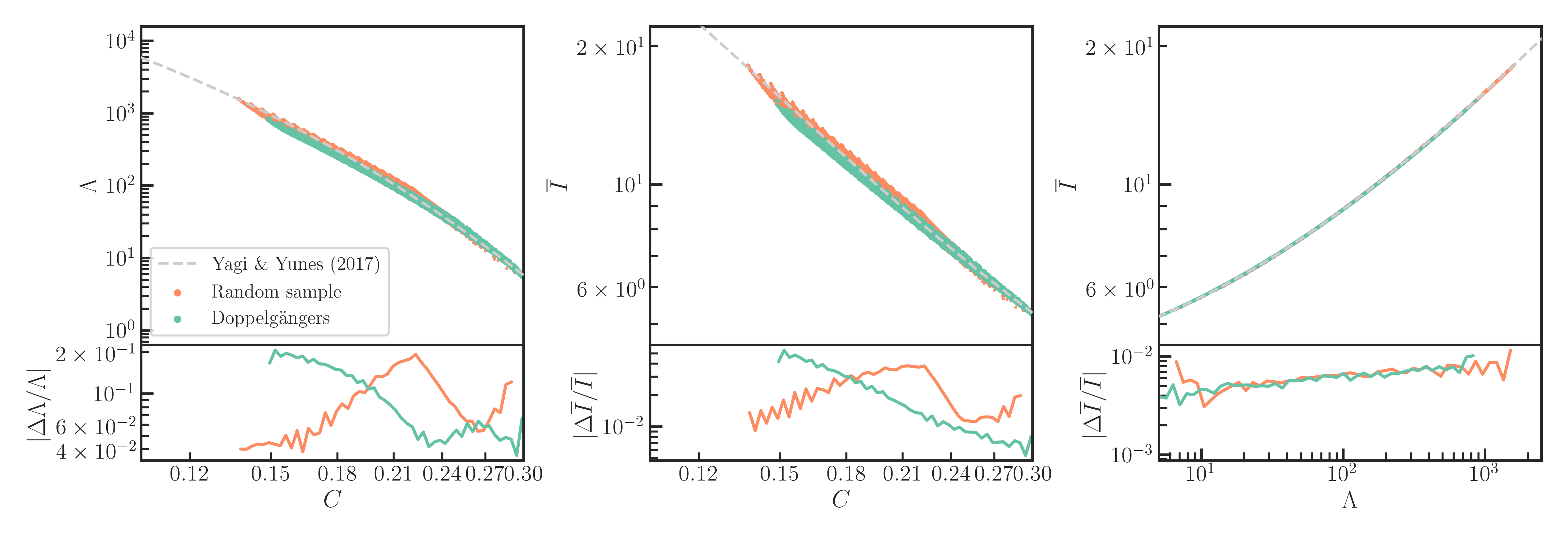}
\caption{\label{fig:ILoveC_15sat} Same as Fig.~\ref{fig:ILoveC},
but for the EoS sample that assumes the crust EoS ap3 to $\rho_0=1.5\rns$,
with the weak prior. Caption details from Fig.~\ref{fig:ILoveC} are repeated for convenience:
Quasi-universal relation between tidal Love number, the neutron star compactness, $C$, and the dimensionless moment of inertia $\overline{I}\equiv I/M^3$.
A random selection of 5,000 EoSs from is shown in orange, while the set of \dops is shown in teal. The quasi-universal relations from \cite{Yagi2017} are shown in gray, dashed lines for reference. The average residuals between the EoS samples and the quasi-universal fit line are shown in the bottom panels.}
\end{figure*}

\begin{figure*}[!ht]
\centering
\includegraphics[width=\textwidth]{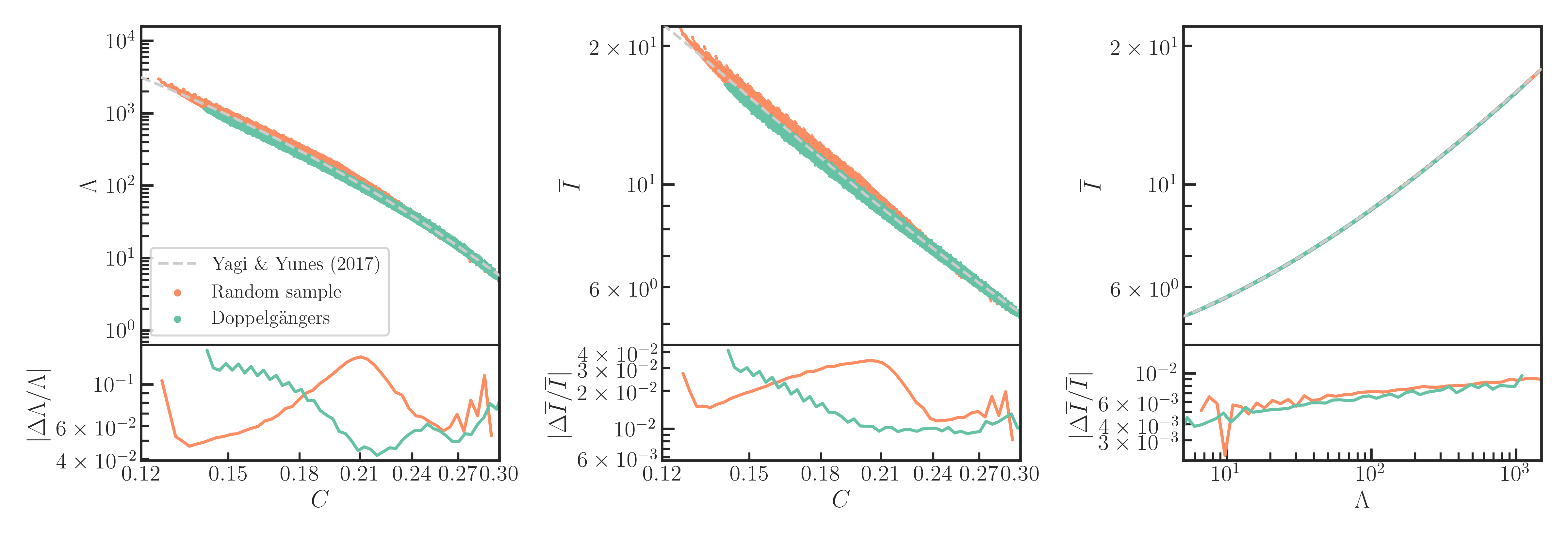}
\caption{\label{fig:ILoveC_reg2} Same as Fig.~\ref{fig:ILoveC_15sat},
but showing the impact of adopting a stronger prior on the pressure derivatives. 
The EoS samples assume the crust EoS ap3 to $\rho_0=\rns$.}
\end{figure*}

\begin{figure*}[!ht]
\centering
\includegraphics[width=\textwidth]{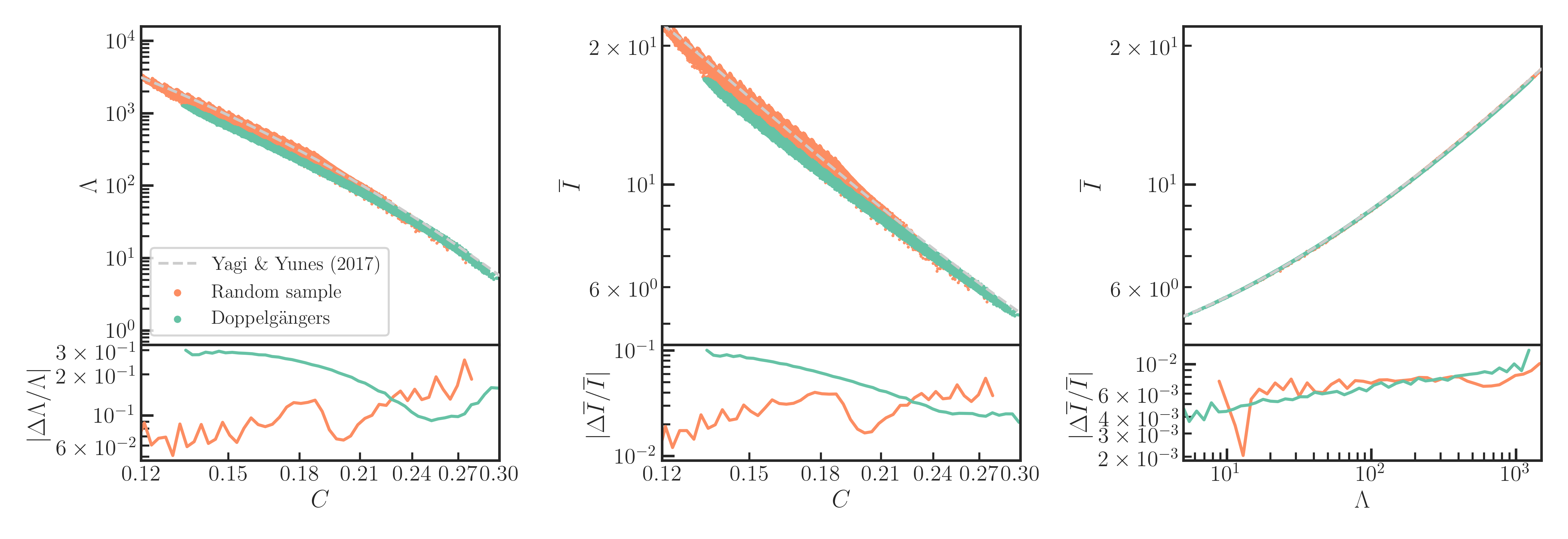}
\caption{\label{fig:ILoveC_ms1b} Same as Fig.~\ref{fig:ILoveC_15sat},
but showing the impact of adopting the stiffer crust EoS, ms1b.  
The EoS samples assume this crust EoS to $\rho_0=\rns$ and adopt a weak prior.}
\end{figure*}

In this appendix, we report the I-Love-C relations
for three additional EoS samples, to demonstrate the impact
of assuming the crust EoS to a higher density of $\rho_0=1.5\rns$
(in Fig.~\ref{fig:ILoveC_15sat}), using a stonger prior
on the high-density pressure derivatives (in Fig.~\ref{fig:ILoveC_reg2}),
and adopting the stiffer crust EoS ms1b (in Fig.~\ref{fig:ILoveC_ms1b}).
In each of these figures, we show the correlations between tidal deformability,
moment of inertia, and compactness for both a set of
randomly-selected EoSs (in orange) or the highest-scoring
set of \dops from a given EoS sample (in teal).

For all three of these EoS samples, which adopt
additional restrictions compared to the baseline EoS sample
described in Sec.~\ref{sec:ILoveC}, we find that the $I-\Lambda$
relation still holds, almost exactly, for the \dop models.
In particular, $I-\Lambda$ correlation for the \dop
models matches that of Ref.~\cite{Yagi:2016bkt} with
sub-percent residuals, indistinguishable from
the correlation for the randomly-selected set of EoS models.

In contrast, we find larger deviations of up to $\sim30\%$
in the $\Lambda-C$ and $I-C$ relations for the \dop models,
compared to the quasi-universal relations of \cite{Yagi:2016bkt}.
The deviations are largest in the $\Lambda-C$ correlations
at small compactness.

Thus, although the EoS samples shown in Figs.~\ref{fig:ILoveC_15sat}-\ref{fig:ILoveC_ms1b} have more restrictions than the baseline EoS sample
of Sec.~\ref{sec:ILoveC}, they deviate more significantly from the existing
quasi-universal $\Lambda-C$ relations. Depending on the region of parameter
space in which the true EoS resides, these deviations may bias
EoS inferences that incorporate the standard $\Lambda-C$ fits; e.g., 
as was done in one of the EoS inferences of the LIGO-Virgo anaylsis of GW170817
\cite{LIGOScientific:2018cki}.

\bibliography{inspire,gwthermal,non_inspire}

\end{document}